\documentclass[fleqn,usenatbib]{mnras}

\usepackage{newtxtext,newtxmath}

\usepackage[T1]{fontenc}
\usepackage{ae,aecompl}
\usepackage{pdflscape}
\usepackage{afterpage}

\usepackage{graphicx}	

\usepackage{amsmath}	
\usepackage{amssymb}	
\usepackage{subfig}
\usepackage{booktabs}
\usepackage{afterpage}
\usepackage{float}
\usepackage{tablefootnote}
\usepackage{hyperref}
\usepackage{longtable}
\usepackage{threeparttable}

\newcommand{\msol}{M_\odot}
\newcommand{\mysim}{\mathord{\sim}}
\newcommand{\myapprox}{\mathord{\approx}}

\usepackage{soul}

\usepackage{isotope} 
\usepackage{float} 
\usepackage{enumitem}
\usepackage{comment}
\newcommand{\nick}{\isotope[56]{Ni} }
\newcommand{\mni}{$M_{\rm{Ni56}}$ }
\newcommand{\mnii}{M_{\rm{Ni56}} }
\newcommand{\mwd}{M_{\rm{WD}} }
\newcommand{\zig}{\tilde{z}_{\rm{ign}} }
\newcommand{\flim}{f_{\rm{lim}} }
\newcommand{\Rig}{R_{\rm{ign}} }
\newcommand{\cmmnt}[1]{}
\newcommand{\mnis}{$M_{\rm{Ni56}}$} 
\defcitealias{kushnir_sub-chandrasekhar-mass_2020}{KWS}
\newcommand{\refKWS}{{\citetalias{kushnir_sub-chandrasekhar-mass_2020}} }
\newcommand{\refKWSs}{{\citetalias{kushnir_sub-chandrasekhar-mass_2020}}}
\title[off-centre ignition]{Off-centre ignition of sub-Chandrasekhar white dwarfs does not resolve the tension with the observed $t_0$-$M_{\rm{Ni}56}$ relation of type Ia supernovae}

\author[Schinasi-Lemberg et al.]{
	Eden Schinasi-Lemberg$^{1,2}$ and Doron Kushnir$^{1}$
	\thanks{E-mail: edenstich@gmail.com} 
	\\
	$^{1}$Dept.of Particle Phys. \& Astrophys., Weizmann Institute of Science, Rehovot 76100, Israel\\
        $^{2}$Dept. of Physics, NRCN, Beer-Sheva 84190, Israel
}

\date{Accepted 2024 December 12; Received 2024 December 9; in original form 2024 October 8}

\pubyear{2023}

\begin{document}
\label{firstpage}
\pagerange{\pageref{firstpage}--\pageref{lastpage}}
\maketitle

\begin{abstract}
Type Ia supernovae (SNe Ia) are likely the thermonuclear explosions of carbon-oxygen (CO) white-dwarf (WD) stars, but the exact nature of their progenitor systems remains uncertain. Recent studies have suggested that a propagating detonation within a thin helium shell surrounding a sub-Chandrasekhar mass CO core can subsequently trigger a detonation within the core (the double-detonation model, DDM). The resulting explosion resembles a central ignition of a sub-Chandrasekhar mass CO WD (SCD), which is known to be in tension with the observed $t_0-$\mni relation, where $t_0$ (the $\gamma$-rays' escape time from the ejecta) is positively correlated with \mni (the synthesized $^{56}$Ni mass). SCD predicts an anti-correlation between $t_0$ and \mnis, with $t_0\myapprox30\,\textrm{day}$ 
for luminous ($\mnii\gtrsim0.5\,M_{\odot}$) SNe Ia, while the observed $t_0$ is in the range of $35-45\,\textrm{day}$. In this study, we apply our recently developed numerical scheme to calculate in 2D the impact of off-centre ignition in sub-Chandrasekhar mass CO WD, aiming to better emulate the behaviour expected in the DDM scenario. Our calculations of the $t_0-\mnii$ relation, which do not require radiation transfer calculations, achieve convergence to within a few percent with a numerical resolution of $\mysim1\,\rm{km}$. We find that the results only slightly depend on the ignition location, mirroring the SCD model, and consequently, the discrepancy with the observed $t_0$-\mni relation remains unresolved.
\end{abstract}





\section{introduction}
\label{sec:introduction}

In recent decades and continuing to the present day, type Ia supernovae (SNe Ia) have played a pivotal role in astrophysical and cosmological research. These supernova events are one of the mechanisms responsible for the synthesis and ejection of elements from silicon to those of the iron peak. Furthermore, thanks to the Phillips relation \citep{phillips_absolute_1993}, they serve as a "standard candle" for assessing cosmological distances. Despite their significant impact, the origins of these events remain a subject of debate. While there is a broad consensus that these events result from the thermonuclear explosion of carbon-oxygen (CO) white dwarfs (WDs), the exact mechanism triggering the explosion remains elusive \citep[for an overview, see][]{maoz_observational_2014}.

One proposed model is known as the "double detonation model" \citep[DDM;][]{nomoto_accreting_1982-1,nomoto_accreting_1982,livne_successive_1990,woosley_sub--chandrasekhar_1994}. In this model, the WD is a component of a binary system and helium is accreted on the WD through Roche-lobe overflow from its companion, leading to the formation of a helium layer on the WD. When this helium layer accumulates sufficient mass or achieves a critical inner density, it initiates a thermonuclear detonation wave (TNDW) within the helium. This detonation can directly ignite the WD core, known as the 'edge-lit' scenario, or it can induce an asymmetric imploding shock, enveloping the WD and converging on an off-centre hotspot where the WD core ignites. While thick helium shells produce too much $^{56}$Ni during nuclear burning for this to be a viable progenitor \citep{hoeflich_explosion_1996,nugent_synthetic_1997,kromer_double-detonation_2010,woosley_sub-chandrasekhar_2011}, recent studies suggested that the minimal mass of a helium shell required to trigger an explosion in the CO core is much smaller than those used in the early models \citep{bildsten_faint_2007,fink_double-detonation_2007,fink_double-detonation_2010,moore_effects_2013,shen_ignition_2014,shen_initiation_2014,shen_almost_2024} and that only minimal amounts of $^{56}$Ni are synthesized in the helium shell, possibly allowing better agreement to observations \citep[see also][]{polin_observational_2019,townsley_double_2019,gronow_sne_2020,boos_multidimensional_2021,gronow_metallicity-dependent_2021,gronow_double_2021,burmester_arepo_2023}. 

Recently, a few tensions between the DDM and observational data have emerged. \citet{SK2021} constructed the intrinsic luminosity function (LF) of SNe Ia using the ZTF Bright Transient Survey catalogue \citep[][]{Fremling2020,Perley2020}. Their analysis revealed unimodal LFs, consistent with previous findings but with significantly lower rates of both dim and luminous events. \citet{ghosh_confronting_2022} showed that for the DDM to explain the low-luminosity suppression derived by \citet{SK2021}, the probability of a low-mass ($\myapprox0.85\,M_{\odot}$) WD explosion should be $\mysim100$-fold lower than that of a high-mass ($\myapprox1.05\,M_{\odot}$) WD. One possible explanation is that the ignition of low-mass CO cores is somehow suppressed, however, \citet{ghosh_confronting_2022} resolved the core ignition in a full-star 1D numerical simulations and showed that if a TNDW can propagate within the He shell, then ignition within the CO core is guaranteed, even for very low-mass, $0.7\,M_{\odot}$, WDs. Other possibilities to explain the low-luminosity suppression include the probability of a WD to be involved in a binary that leads to the required conditions for the DDM to operate is much lower for low-mass WDs than for high-mass WDs \citep[although binary population synthesis calculations do not support this, see][]{Shen2017}. Alternatively, it could be that the ignition probability of a TNDW within the Helium shell is suppressed for low-mass WDs. While the ignition mechanism of the Helium shell is not fully understood \citep[see, e.g.,][]{Zingale2013,Jacobs2016,Glasner2018}, one could speculate that since low-mass WDs have a lower virial temperature, then more stringent conditions are required from the progenitor binary to achieve ignition. Finally, achieving a sufficiently high density for a propagating TNDW may require more massive helium shells for low-mass WDs, potentially excluding certain binary configurations \citep[see, e.g.,][]{piersanti_expected_2024}. 

\begin{figure*}
    	\includegraphics[width=15cm]{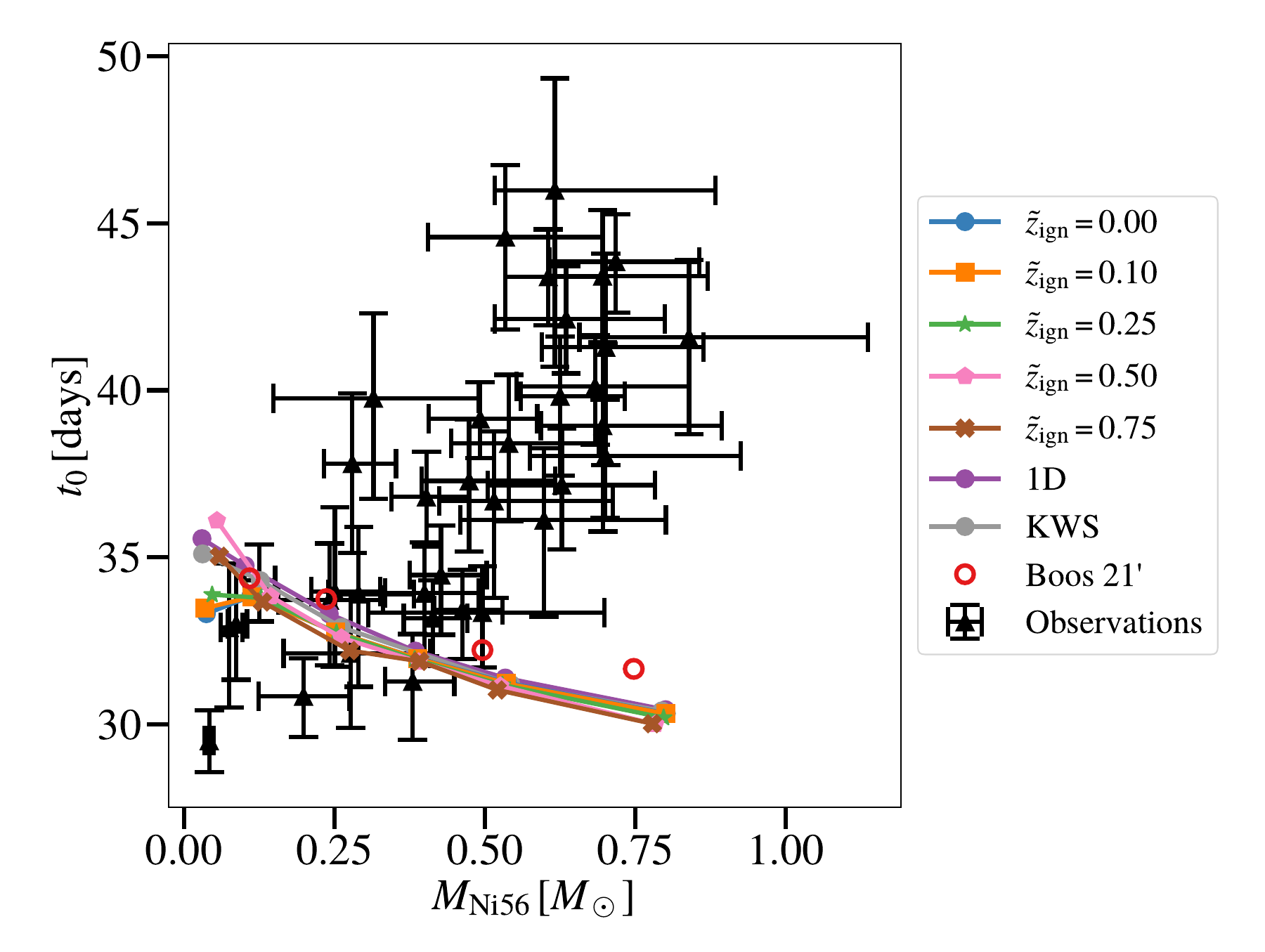}
    \caption{The $t_0-M_{\rm{Mi56}}$ relation. Black circles: the observed SNe Ia sample \protect\citep{sharon_-ray_2020,sharon_personal_2023}. Purple line: 1D calculation of our setup (see description in Section~\ref{sec:setup}) for $\mwd=0.8,0.85,0.9,0.95,1.0,1.1\msol$ (left to right). Grey line: the 1D SCD results of \refKWSs. Our setup shows minor deviations from \refKWS results. There is a clear tension between SCD predictions and the observed $t_0-$\mni relation. SCD predicts an anti-correlation between $t_0$ and \mnis, with $t_0\myapprox30\,\textrm{day}$ for luminous (\mni$\gtrsim0.5\,M_{\odot}$) SNe Ia, while the observed $t_0$ is in the range of $35-45\,\textrm{day}$. Different colour symbols (blue, orange, green, pink, and brown): results of 2D calculations with the same dimensionless ignition point $\zig = z_{\rm{ign}}/R_{\rm{WD}}$ (0, 0.1, 0.25, 0.5, 0.75, respectively), where $z_{\rm{ign}}$ is the ignition location and $R_{\rm{WD}}$ is the WD radius. Each series includes calculations with $\mwd=0.8,0.85,0.9,0.95,1.0,1.1\msol$ (left to right). Our 2D results show only slight dependence on ignition location, mirroring the 1D calculations of \refKWSs, leaving the discrepancy with the observed $t_0$-\mni relation unresolved. Red circles: results from the DDM simulations conducted by \citet[][]{boos_multidimensional_2021}, as presented in \citet[][]{boos_type_2024}.}
    \label{fig:main_plot}
\end{figure*}

Arguably the most prominent discrepancy with observations concerns the optical depth of $\gamma$-rays resulting from the decay of radionuclides synthesized during the explosion. The $\gamma$-ray optical depth is high shortly following the explosion, so all the $\gamma$-ray energy is deposited within the ejecta. As the ejecta expands, the $\gamma$-ray optical depth decreases, and some $\gamma$-rays only partially deposit their energy or escape the ejecta without interacting \citep{jeffery_radioactive_1999}. The $\gamma$-ray escape time, $t_0$ \citep{Stritzinger2006,Scalzo2014,wygoda_type_2019}, is defined by \citep{jeffery_radioactive_1999}
\begin{equation}\label{eq:t0eq}
f_\text{dep}(t) = \frac{t_0^2}{t^2},\;\;\;t\gg t_0\; (f_\text{dep}\ll 1),
\end{equation}
where $t$ is the time since the explosion and $f_\text{dep}(t)$ is the $\gamma$-ray deposition function, which describes the fraction of the generated $\gamma$-ray energy that is deposited in the ejecta. For sufficiently small $\gamma$-ray optical depth, the deposition function is proportional to the column density, which scales as $t^{-2}$. The value of $t_0$ can be measured from a bolometric light curve with a few percent accuracy \citep[for typical available observational data; see][]{wygoda_type_2019} due to an integral relation derived by \cite{katz_exact_2013}, independent of the supernova distance. Together with the $^{56}$Ni mass synthesized in the explosion, \mnis, an observed $t_0-$\mni relation can be constructed \citep{wygoda_type_2019}. \citet{sharon_-ray_2020} have accurately determined $t_0$ to discover a positive correlation between $t_0$ and \mnis, see Fig.~\ref{fig:main_plot}. The observed $t_0-$\mni correlation is similar to the Phillips relation \citep{phillips_absolute_1993}, which relates the maximum flux to the width of the light curve in some bands. However, unlike the Phillips relation, comparing models to the $t_0-$\mni relation bypasses the need for radiation transfer calculations, as the value of $t_0$ can be directly inferred from the ejecta (at least in cases where the deviation from spherical symmetry is not significant, see below).

\citet[][hereafter \refKWSs]{kushnir_sub-chandrasekhar-mass_2020} performed 1D calculations of centrally-ignited solar-metallicity sub-Chandrasekhar WDs (SCD), employing a modified version of the Eulerian, hydrodynamic \textsc{FLASH4.0} code \citep{fryxell_flash_2000,dubey_extensible_2009} with a 178 isotopes-list. They showed that the calculated \mni and $t_0$ converge to an accuracy better than a few percent. The converged results of these calculations are presented in Fig.~\ref{fig:main_plot}. As can be seen in the figure, there is a clear tension between the predictions of SCD and the observed $t_0-$\mni relation. SCD predicts an anti-correlation between $t_0$ and \mnis, with $t_0\myapprox30\,\textrm{day}$ for luminous (\mni$\gtrsim0.5\,M_{\odot}$) SNe Ia, while the observed $t_0$ is in the range of $35-45\,\textrm{day}$. They showed that various uncertainties related to the physical processes and the initial profiles of the WD are unlikely to resolve the tension with the observations. At the same time, they can reduce the agreement with the observations for low-luminosity SNe Ia. 

\refKWS proposed that the discrepancies between the 1D SCD model and observational data might be mitigated by conducting more intricate simulations based on the more realistic DDM. While the nucleosynthesis and the energy release within the thin He layer are unlikely to significantly affect either \mni or $t_0$, other differences between SCD and DDM could influence the simulation outcomes:
\begin{enumerate}[wide, labelwidth=!,itemindent=!,labelindent=0pt, leftmargin=0em, parsep=0pt]
\item The different initial conditions of the CO core due to the compression wave that propagates in the CO core prior to ignition.
\item The asymmetrical dynamics resulting from off-centre ignition of the CO core.
\item The interaction of the ejecta with the companion star.
\end{enumerate}
Studying these effects necessitates comprehensive multidimensional simulations that exceed the scope of this paper. Here, we specifically investigate the impact of off-centre ignition of CO WD using 2D hydrodynamical simulations. We demonstrate that we can achieve convergence in the calculated \mni and $t_0$, thereby showing that this effect does not resolve the tension with the observed $t_0-$\mni relation.

A multi-dimensional structure of the ejecta poses challenges in interpreting $t_0$, which is derived observationally from the bolometric light curve, subject to viewing-angle dependence. This dependency can be mitigated by late times ($\gtrsim$1 year) observations, which would provide the average $t_0$ of the SN irrespective of viewing angle \citep[for a detailed discussion, see][]{sharon_all_2024}. However, at these late phases, a significant fraction of the flux falls within the MIR range \citep{Chen2023}, requiring multi-epoch \textit{JWST} observations, feasible for only a few objects. To interpret observations at $\mysim100\,\textrm{d}$ \citep[typically required to determine $t_0$;][]{sharon_-ray_2020}, a calculation of the bolometric light curve (including the viewing angle dependence) is needed. For such calculations, the ejecta plasma can no longer be assumed to be in local thermodynamic equilibrium (LTE; the LTE approximation is valid up to $\mysim30\,\rm{d}$), significantly complicating calculations. Currently, there is no numerical radiation transfer code for multi-dimensional ejecta capable of calculating the bolometric light curve without LTE assumptions. In this paper, we only compute the average $t_0$ of the obtained ejecta, which can be efficiently estimated using an analytical expression. Since the ejecta in our study displays minimal asymmetries and the deviations of the average $t_0$ values from centrally-ignited values are small, it is reasonable to rely on the average $t_0$ approximation. We postpone a more thorough investigation of the viewing-angle effect to future studies.

In Section \ref{sec:setup}, we outline the setup of our calculations, which is based on the configuration utilized by \refKWS and employs a modified version of the \textsc{FLASH4.0} code. Our setup has been adjusted to enable off-centre ignition calculations in a 2D simulation while remaining within reasonable computational resource constraints. 1D simulations with a setup resembling our 2D configuration exhibit minor deviations from the 1D results of \refKWSs, as shown in Fig.~\ref{fig:main_plot}.

Section \ref{sec:example} delves into the dynamic evolution of the simulations, elucidating the specifics through two particular cases. This section also discusses the convergence of the simulations and investigates the impact of various aspects of the numerical configuration on the results.

In Section \ref{sec:results}, we present the outcomes of our entire simulation suite. The key finding of this investigation is summarized in Fig.~\ref{fig:main_plot}. Our results indicate only a slight reliance on the ignition location, mirroring the 1D calculations of \refKWSs, and consequently, the discrepancy with the observed $t_0$-\mni relation remains unresolved. Finally, Section \ref{sec:summary} provides a summary of our findings and a discussion of future research directions. All data of the ejecta used to derive the presented results is available in \url{https://drive.google.com/drive/folders/1RzQzcvCYmVnN_Eq8oiZtNd8kJ9kMhvyQ?usp=sharing}.


\section{Calculation setup}
\label{sec:setup}

This section provides an overview of our computational setup, primarily based on the configuration utilized by \refKWSs. Section~\ref{sec:flash} introduces the modified version of the \textsc{FLASH4.0} code employed in our study. Additional adjustments to the code and the numerical scheme, relative to those described in \refKWSs, are presented in Section~\ref{sec:modif}. Sections~\ref{sec:init} and~\ref{sec:igni} delve into the initial setup and the ignition methodology, respectively. Our approach to estimating the properties of the freely-expanding ejecta is detailed in Section~\ref{sec:ejecta}.

\subsection{Modified \textsc{FLASH} code}
\label{sec:flash}

We use the \textsc{FLASH4.0} code, modified to incorporate thermonuclear reaction rates computed using the \textsc{MESA} code \citep{2011ApJS..192....3P} and the input physics of \refKWSs. Additionally, our code features the burning scheme of  \cite{kushnir_accurate_2020}, which enhances the accuracy of TNDW calculations while maintaining reasonable computational efficiency. The two key components of this burning scheme are as follows:
\begin{enumerate}[wide, labelwidth=!,itemindent=!,labelindent=0pt, leftmargin=0em, parsep=0pt]
\item A burning limiter that constrains the rates of thermonuclear reactions to ensure that the typical time-scale for changes, such as energy release, within a computational cell is limited to a fraction ($\flim$) of the sound-crossing time of the cell. This adjustment results in the smearing of the TNDW burning zone, which in reality is significantly smaller compared to the resolution, over approximately $1/\flim$ computational cells, while preserving the thermodynamic trajectory along the burning zone. As demonstrated in \cite{kushnir_accurate_2020}, this approach enables an accurate calculation of the TNDW structure. Throughout this paper, unless otherwise specified, we set the burning limiter for the time-scales of energy release and change of $\Tilde{Y}=\sum\limits_{i\neq n,p,\alpha}Y_i$ to $\flim=0.1$. Here, $Y_i$ represents the molar fraction of the $i$-th isotope, which can be approximated as $Y_i\myapprox X_i/A_i$, with $X_i$ and $A_i$ denoting the mass fraction and nucleon number, respectively. 
\item An adaptive separation of isotopes into groups in quasi-nuclear-statistical equilibrium that resolves the time-consuming burning calculation of reactions that are nearly balanced out. Burning is calculated \textit{in situ}, employing the required large networks without post-processing or pre-describing the conditions behind the TNDW.
\end{enumerate}

\subsection[Modification to KWS]{Modifications to \refKWS}
\label{sec:modif}
Several adjustments were implemented to the 1D calculations presented by \refKWS to facilitate 2D simulations within reasonable computational timeframes. These adjustments include a modified adaptive mesh refinement (AMR) scheme (see Appendix~\ref{sec:amr}) and a reduced list of 38 isotopes instead of the 178 (or 69) isotopes utilized by \refKWSs. The 38-isotopes list include: n, p, $\alpha$, \isotope[12]{C}, \isotope[13]{N}, \isotope[16, 17]{O}, \isotope[17]{F}, \isotope[20, 22]{Ne}, \isotope[23]{Na}, \isotope[24, 26]{Mg}, \isotope[27]{Al}, \isotope[28-30]{Si}, \isotope[32, 34]{S}, \isotope[35]{Cl}, \isotope[36, 38]{Ar}, \isotope[39]{K}, \isotope[40, 42]{Ca}, \isotope[45]{Sc}, \isotope[44-46]{Ti}, \isotope[47, 48]{V}, \isotope[48-50]{Cr}, \isotope[53]{Mn}, \isotope[54]{Fe}, \isotope[55]{Co}, and \isotope[56]{Ni}. 

A crucial adaptation made to simulate the off-centre ignition involves subdividing the calculation grid into 50 discrete angles around the ignition point. At each angle, we calculate the outgoing shock radius and its breakout time from the surface of the star, both of which are essential for the AMR scheme (see Appendix~\ref{sec:amr}). 

The impact of these modifications is relatively minor and is elaborated upon in Appendix~\ref{sec:numericsens}.

\subsection{Initial setup}
\label{sec:init}

The initial profile is generated using a similar procedure as described in \refKWSs. Specifically, we employ a modified routine for calculating cold WD structure, developed by Frank Timmes\footnote{\url{https://cococubed.com/code_pages/coldwd.shtml}} to calculate the hydrostatic equilibrium of a solar-metallicity CO WD. The WDs have a uniform composition, representing Solar metallicity: $X_{\rm{C12}}=X_{\rm{O16}}=0.4925$ and $X_{\rm{Ne22}}=0.015$, and are isothermal with $T=10^7\,\rm{K}$. The WD profile is then interpolated to the computational mesh. The external region surrounding the WD is assigned a density and temperature of \texttt{amb\_dens}=$0.01\,\rm{g/cm^3}$ and \texttt{amb\_temp}=$10^7\,\rm{K}$, respectively. To mitigate the impact of numerical deviations from hydrostatic equilibrium in unperturbed regions, we enforce the initial state for each cell within these areas. 

The computational mesh extends in the radial distance $R=[0,L]$ and altitude $Z=[-L,L]$, where $L=2^{17}\,\rm{km}\myapprox1.31\times10^5\,\rm{km}$ ($L\myapprox1.31\times10^6\,\rm{km}$ in some cases). Boundary conditions include a solid wall on the lower radial axis (symmetry axis) and free surfaces on all other boundaries. Our base mesh consists of $4\times4$ cell blocks with 4 blocks in the $R$-direction and 8 blocks in the $Z$-direction (40 and 80 for $L\myapprox1.31\times10^6\,\rm{km}$). For $\mwd>0.9\msol$, the maximal refinement level considered was 14, which corresponds to a maximal resolution of $\Delta x_{\rm{min}}=1\,\rm{km}$. For cases of $\mwd\le0.9\msol$, we have chosen to use a maximal refinement level of 15 ($\Delta x_{\rm{min}}=0.5\,\rm{km}$), as the results tend to require more resolution to converge, due to the lower amount of synthesized \nick.

\subsection{Ignition method}
\label{sec:igni}

The ignition method we employ replicates that presented in \refKWSs. This method involves setting the temperature to $4\times10^9\,\rm{K}$ for $r \leq \Rig$ and configuring the composition within this region to match the nuclear-statistical-equilibrium (NSE) state. Within this region, the radial velocity follows a linear profile in the range [0, $\Rig$], with $v(0)=0$ and $v(\Rig)=2\times10^4\,\rm{km/s}$. The radius is measured relative to the centre of the ignition region $z_{\rm{ign}}$, not necessarily located at the centre of the WD. To accommodate different WD masses, we introduce the dimensionless ignition location, $\zig = z_{\rm{ign}}/R_{\rm{WD}}$, where $R_{\rm{WD}}$ is the radius of the WD. When adjusting the maximum resolution or $\flim$, the radius of the ignition region changes according to $ \Rig \propto \Delta x_{\rm{min}}/\flim $, with fiducial values of $ \Rig=100(200)\,\rm{km} $ for $\Delta x_{\rm{min}} = 1\,\rm{km}$ and $\flim=0.1$ with ignition locations of $\zig\le0.5(=0.75)$. 

The chosen values of $\Rig$ are larger by a factor of 2(4) from the values used in the low-mass (high-mass) WD calculations of \refKWS (compared at the same resolution and limiter). This ensures that off-centre ignitions (at densities below the central density of the WD) achieve a stable TNDW. When we conduct 2D simulations with central ignitions using a similar $\Rig$ as chosen by \refKWSs, we observe a minimal impact on our results, with deviations of less than $1\,\%$ compared to calculations with central ignition using our $\Rig$ values. In cases where $\Rig$ proves inadequate for a successful ignition, mainly in low-mass WD simulations, we increase it by a factor of 2 or 4. The parameters for each simulation in this work are given in Table~\ref{tab:tabres}.

\begin{table*}
\caption{The obtained $\mnii$ (seventh column) and $t_0$ (eigth column) as a function
of $\mwd$ (first column) and  $\zig$ (second column). Other simulation parameters [$\Rig$, $\Delta x_{\min}$, the number of isotopes ($N_{\rm{iso}}$), and $\flim$] are provided in the third to sixth columns, respectively. The entire table is available in electronic form (\url{https://drive.google.com/drive/folders/1RzQzcvCYmVnN_Eq8oiZtNd8kJ9kMhvyQ?usp=sharing}).} \label{tab:tabres}
\begin{tabular}{cccccccc} 
\hline
$\mwd[\msol]$ & $\zig$ & $\Rig[\rm{km}]$ & $ \Delta x_{\min}[\rm{km}]$ & $N_{\rm{iso}}$ & $\flim$ & $\mnii[\msol]$ & $t_0[\rm{days}]$ \\ \hline
$0.80$ & $0.00$ & $50$ & $0.5$ & $38$ & $0.10$ & $3.73 \times 10^{-2}$ & $33.3$ \\
$0.85$ & $0.00$ & $50$ & $0.5$ & $38$ & $0.10$ & $1.11 \times 10^{-1}$ & $33.8$ \\
$0.90$ & $0.00$ & $50$ & $0.5$ & $38$ & $0.10$ & $2.52 \times 10^{-1}$ & $32.8$ \\
$0.95$ & $0.00$ & $100$ & $1.0$ & $38$ & $0.10$ & $3.89 \times 10^{-1}$ & $32.0$ \\
$1.00$ & $0.00$ & $100$ & $1.0$ & $38$ & $0.10$ & $5.37 \times 10^{-1}$ & $31.2$ \\
$1.10$ & $0.00$ & $100$ & $1.0$ & $38$ & $0.10$ & $8.01 \times 10^{-1}$ & $30.3$ \\
$0.80$ & $0.10$ & $50$ & $0.5$ & $38$ & $0.10$ & $3.43 \times 10^{-2}$ & $33.5$ \\
$0.85$ & $0.10$ & $50$ & $0.5$ & $38$ & $0.10$ & $1.13 \times 10^{-1}$ & $33.8$ \\
$0.90$ & $0.10$ & $50$ & $0.5$ & $38$ & $0.10$ & $2.52 \times 10^{-1}$ & $32.8$ \\
$0.95$ & $0.10$ & $100$ & $1.0$ & $38$ & $0.10$ & $3.88 \times 10^{-1}$ & $32.0$ \\
$1.00$ & $0.10$ & $100$ & $1.0$ & $38$ & $0.10$ & $5.36 \times 10^{-1}$ & $31.2$ \\
$1.10$ & $0.10$ & $100$ & $1.0$ & $38$ & $0.10$ & $8.01 \times 10^{-1}$ & $30.3$ \\
$0.80$ & $0.25$ & $50$ & $0.5$ & $38$ & $0.10$ & $4.69 \times 10^{-2}$ & $33.9$ \\
$0.85$ & $0.25$ & $50$ & $0.5$ & $38$ & $0.10$ & $1.28 \times 10^{-1}$ & $33.8$ \\
$0.90$ & $0.25$ & $50$ & $0.5$ & $38$ & $0.10$ & $2.53 \times 10^{-1}$ & $32.7$ \\
$0.95$ & $0.25$ & $100$ & $1.0$ & $38$ & $0.10$ & $3.83 \times 10^{-1}$ & $31.9$ \\
$1.00$ & $0.25$ & $100$ & $1.0$ & $38$ & $0.10$ & $5.29 \times 10^{-1}$ & $31.2$ \\
$1.10$ & $0.25$ & $100$ & $1.0$ & $38$ & $0.10$ & $7.97 \times 10^{-1}$ & $30.2$ \\
$0.80$ & $0.50$ & $200$ & $0.5$ & $38$ & $0.10$ & $5.49 \times 10^{-2}$ & $36.1$ \\
$0.85$ & $0.50$ & $100$ & $0.5$ & $38$ & $0.10$ & $1.44 \times 10^{-1}$ & $33.8$ \\
$0.90$ & $0.50$ & $50$ & $0.5$ & $38$ & $0.10$ & $2.62 \times 10^{-1}$ & $32.6$ \\
$0.95$ & $0.50$ & $100$ & $1.0$ & $38$ & $0.10$ & $3.90 \times 10^{-1}$ & $31.9$ \\
$1.00$ & $0.50$ & $100$ & $1.0$ & $38$ & $0.10$ & $5.26 \times 10^{-1}$ & $31.1$ \\
$1.10$ & $0.50$ & $100$ & $1.0$ & $38$ & $0.10$ & $7.83 \times 10^{-1}$ & $30.0$ \\
$0.80$ & $0.75$ & $400$ & $0.5$ & $38$ & $0.10$ & $5.89 \times 10^{-2}$ & $35.0$ \\
$0.85$ & $0.75$ & $400$ & $0.5$ & $38$ & $0.10$ & $1.31 \times 10^{-1}$ & $33.7$ \\
$0.90$ & $0.75$ & $400$ & $0.5$ & $38$ & $0.10$ & $2.75 \times 10^{-1}$ & $32.2$ \\
$0.95$ & $0.75$ & $400$ & $1.0$ & $38$ & $0.10$ & $3.92 \times 10^{-1}$ & $31.9$ \\
$1.00$ & $0.75$ & $200$ & $1.0$ & $38$ & $0.10$ & $5.21 \times 10^{-1}$ & $31.0$ \\
$1.10$ & $0.75$ & $200$ & $1.0$ & $38$ & $0.10$ & $7.78 \times 10^{-1}$ & $30.0$ \\
\hline
\centering
\end{tabular}
\end{table*}

\subsection{Ejecta properties}
\label{sec:ejecta}

We examine throughout the simulation the total kinetic energy, $E_{\rm{kin}}$, the total internal energy, $E_{\rm{int}}$, and the total gravitational energy, $E_{\rm{grav}}$. We stop the simulation when both $E_{\rm{kin}}/E_{\rm{int}}>20$ and $-E_{\rm{kin}}/E_{\rm{grav}}>20$ (typically the former condition is fulfilled later). At this point, deviations from homologous expansion are typically within a few percent. The velocity of each cell, $v_i$, for the asymptotic freely expanding ejecta, is determined by $v_i=r_i/t_{\rm{eff}}$, where $r_i$ is the radius of each cell, and $t_{\rm{eff}}$ is determined such that the total kinetic energy of the asymptotic ejecta equals $E_{\rm{kin}}$. 

From the asymptotic freely expanding ejecta of each calculation, we determine $t_0$ using the following expression \citep{wygoda_type_2019}:
\begin{equation}\label{eq:sigmav}
    {t_0^2=\frac{\kappa_{\rm{eff}}}{M_{\rm{Ni56}}}\int^{\infty}_{0}dvv^2\rho(v)X_{\rm{Ni56}}(v)\int d\hat{\Omega}\int^{\infty}_{0}ds\rho(\vec{v}+s\hat{\Omega})},
\end{equation}
where $\rho(v)$ represents the mass density, $X_{\rm{Ni56}}(v)$ represents the mass fraction of $^{56}$Ni, $\kappa_{\rm{eff}}\myapprox0.025(Y_e/0.5)\,\rm{cm^2g^{-1}}$ \citep{swartz_gamma-ray_1995,jeffery_radioactive_1999} is the effective opacity, and we use $Y_e=0.5$. To estimate the uncertainty associated with the evaluation of $t_0$, we perform Monte-Carlo (MC) $\gamma$-ray transport calculations to determine $f_{\rm{dep}}$ for the converged asymptotic freely expanding ejecta, employing the methods outlined in \citet{sharon_-ray_2020}. At late times, we get $f_{\rm{dep}}\propto t^{-2}$, allowing us to determine $t_{0}^{\gamma\rm{RT}}=f_{\rm{dep}}^{1/2}t$. For all inspected cases, the deviation of $t_{0}^{\gamma\rm{RT}}$ from $t_0$ is $\lesssim5\%$ (and for most of them $<1\%$).


\section{Example calculations}
\label{sec:example} 
This section explores the dynamics and outcomes of two simulation examples with $\zig=0.5$. The first instance, featuring $\mwd=0.8\,\msol$, is detailed in Section \ref{sec:ex08}, while the second example, involving $\mwd=1.1\,\msol$, is covered in Section \ref{sec:ex11}. A convergence study (also considering the impact of the burning limiter) of these examples, along with analogous calculations employing central ignition ($\zig=0$), is presented in Section~\ref{sec:exconv}. The sensitivity to the isotope list is discussed as well.

\subsection[Example 1]{Example 1: $\mwd=0.8\,\msol$ and $ \zig=0.5 $ }
\label{sec:ex08}
\begin{figure*}
\includegraphics[height=7cm]{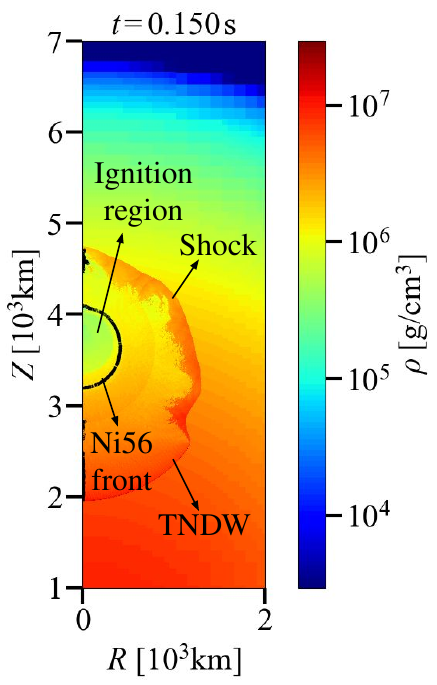}{(a)}
\includegraphics[height=7cm]{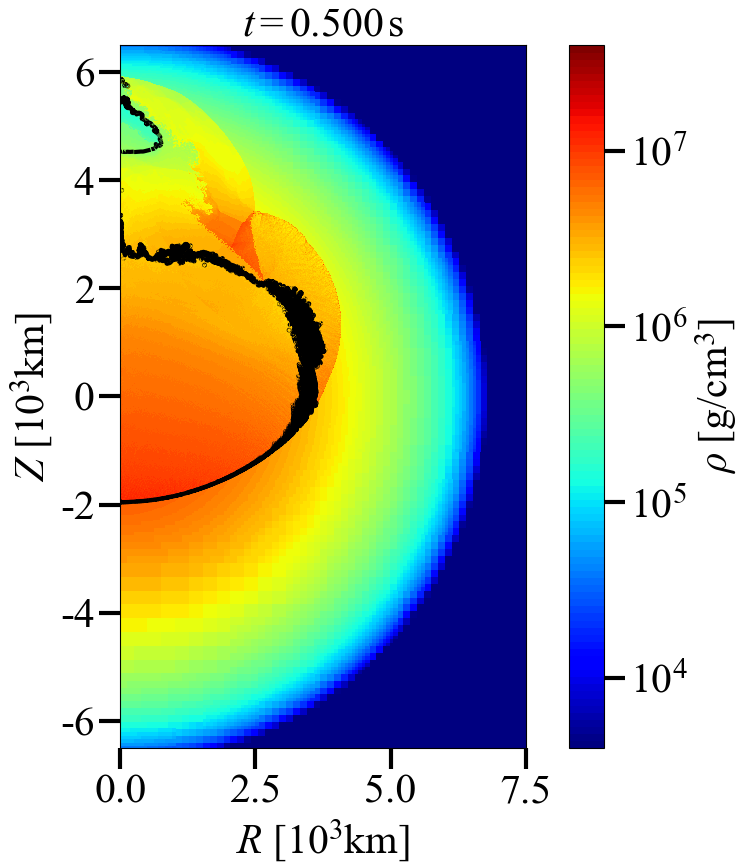}{(b)} \\
\includegraphics[height=7cm]{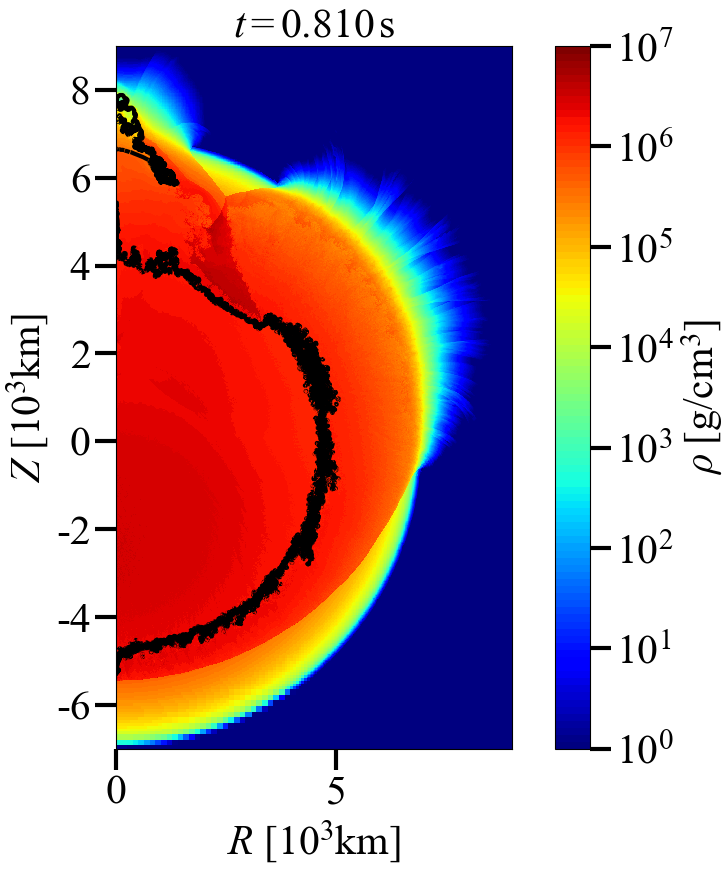}{(c)}
\includegraphics[height=7cm]{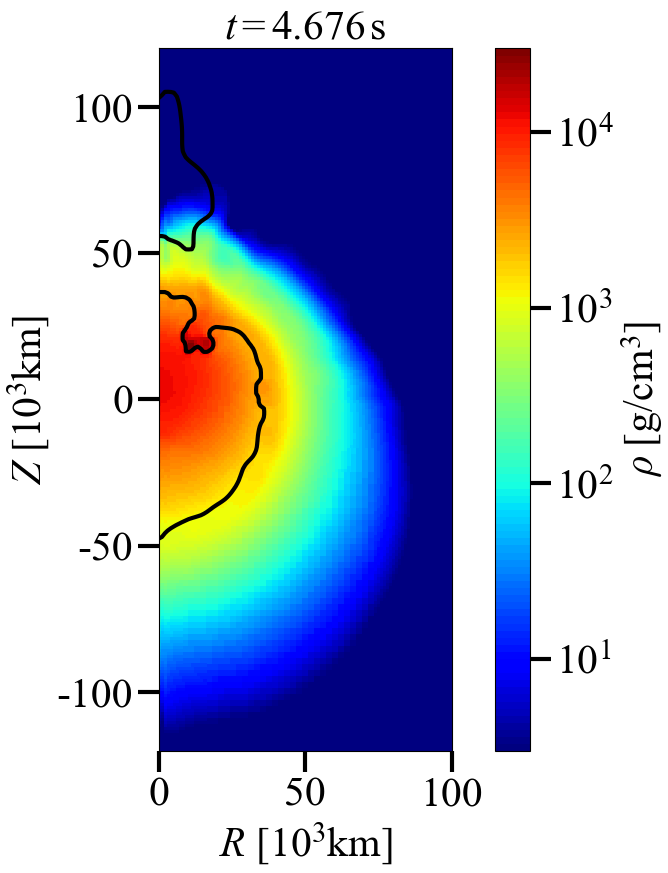}{(d)}
\caption{Dynamical evolution of the explosion of a $0.8\,\msol$ WD with $\zig=0.5$ ($\Delta x_{\rm{min}}=0.5\,\rm{km}$ and $\Rig=200\,\rm{km}$). The colour map represents the density, and the black contour represents $X_{\rm{Ni56}}=0.01$. Panel (a): due to low densities in the ignition region, $^{56}$Ni is not synthesized immediately behind the TNDW, except in the initially ignited region. Panel (b): as the TNDW encounters even lower densities (large $Z$ values, north), it transforms into a regular shock. In the opposite direction (low $Z$ values, south), the TNDW encounters high densities of $\mysim10^7\,\rm{g/cm^3}$, facilitating efficient burning and producing $^{56}$Ni. Panel (c): the asymmetric arrival of the TNDW at the edge of the WD. The breakout time varies non-monotonically with the distance from the ignition point due to differences in shock speeds between the north and south directions, arising from varying upstream densities. Panel (d): the mass distribution during the homologous expansion stage. The final $^{56}$Ni distribution comprises two regions. The main central region is slightly shifted to the south and forms as the TNDW moves through regions of higher densities than the ignition point. The second, northern region results from the TNDW immediately after ignition and may be influenced by the details of the numerical ignition process.}
\label{fig:m08}
\end{figure*}

Fig.~\ref{fig:m08} illustrates the dynamic evolution of $\mwd=0.8\,\msol$ with $ \zig=0.5$ ($\Delta x_{\rm{min}}=0.5\,\rm{km}$ and $\Rig=200\,\rm{km}$). Due to the low densities in the ignition region, \nick is not synthesized immediately behind the TNDW, except for the region initially ignited [panel (a); The black contour on each panel represents the value $ X_{\rm{Ni}}=0.01$]. As the TNDW encounters even lower densities (large $Z$ values, north), it transforms into a regular shock. In the opposite direction (low $Z$ values, south), the TNDW encounters high densities of $\mysim10^7\,\rm{g/cm^3}$, facilitating efficient burning and resulting in the production of \nick (panel (b)).

Panel (c) illustrates the asymmetric arrival of the TNDW at the edge of the WD. The breakout time varies non-monotonically with the distance from the ignition point due to differences in shock speeds between the north and south directions, arising from varying upstream densities. Additionally, the outer equatorial region of the ejecta appears segmented angularly. Given that the number of these segments exceeds our angular discretization of the WD (see Section~\ref{sec:setup}), this effect likely originates from TNDW instability, elaborated on in Appendix~\ref{sec:detinstab}.

Panel (d) displays the mass distribution during the homologous expansion stage. The final distribution of \nick consists of two regions. The main central region is slightly shifted to the south and forms as the TNDW moves through regions of higher densities than the ignition point. The second, northern region, results from the TNDW immediately after ignition and may be influenced by the details of the numerical ignition process. The northern region is in a low-density area and has a negligible effect on both $\mnii$ and $t_0$.

This example yields $ \mnii=0.055\,\msol $ and $ t_0=36.1 $ d, while the 2D simulation of the symmetric case produces $ \mnii = 0.037\,\msol $ and $t_0=33.3$ d, slightly diminishing the agreement with observations, as depicted in Fig.~\ref{fig:main_plot}. It is important to note that we compare the results with a 2D simulation of the symmetrical case rather than the 1D simulation, as the latter fails to capture the 2D TNDW instability, impacting the outcomes (see Appendix~\ref{sec:detinstab}).

\subsection[Example 2]{Example 2: $\mwd=1.1\,\msol$ and $ \zig=0.5 $ }
\label{sec:ex11}
\begin{figure*}
\includegraphics[height=7cm]{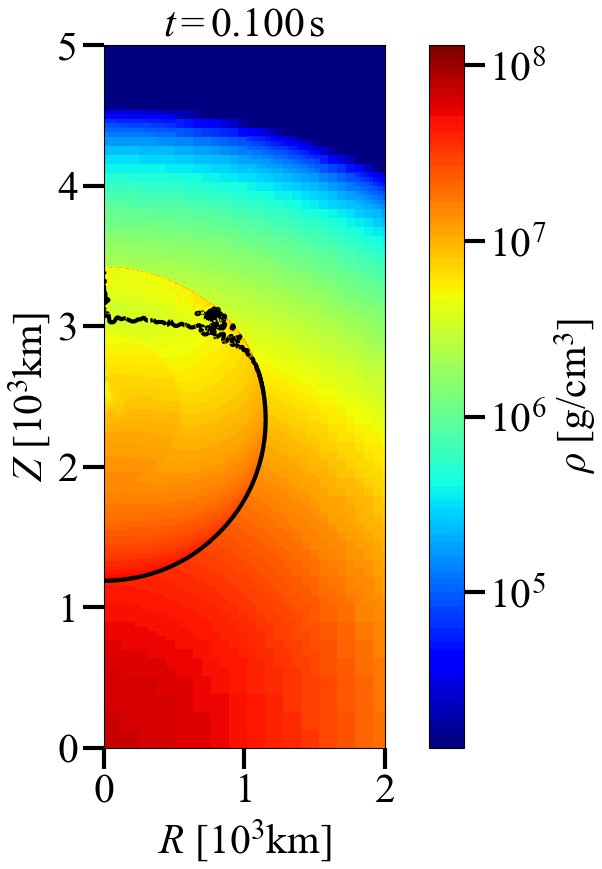}{(a)}
\includegraphics[height=7cm]{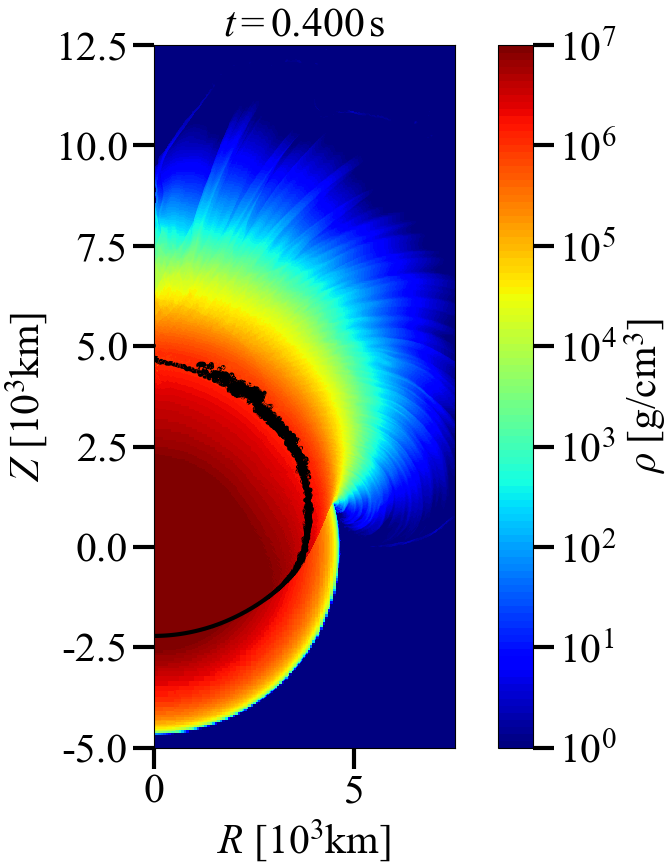}{(b)}
\includegraphics[height=7cm]{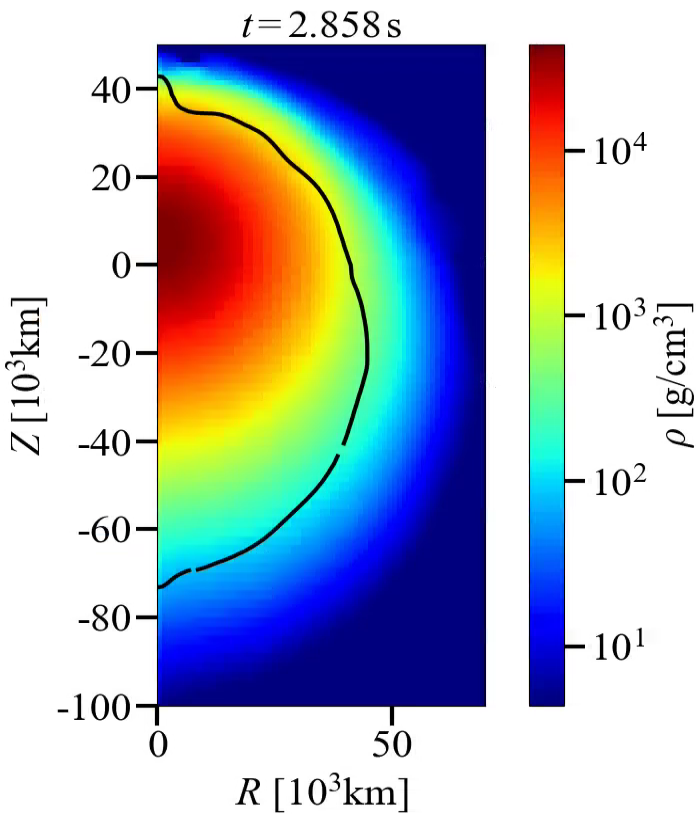}{(c)}
\caption{Dynamical evolution of the explosion of a $1.1\,\msol$ WD with $\zig=0.5$ ($\Delta x_{\rm{min}}=1\,\rm{km}$ and $\Rig=100\,\rm{km}$). The colour map represents the density, and the black contour represents $X_{\rm{Ni56}}=0.01$. Panel (a): due to the higher density around the ignition zone, $^{56}$Ni is consistently produced behind the TNDW front immediately after ignition. Consequently, the $^{56}$Ni front appears almost spherical, except for the northernmost region where the TNDW encounters lower densities. Panel (b): the structure around the breakout time on the northern side of the WD. Panel (c): the ejecta at the homologous expansion stage. Unlike the $0.8\,\msol$ WD case, no significant region in the ejecta is associated with the early ignition stage. The ejecta appears more spherical, and the $^{56}$Ni mass is distributed in a single region.}
\label{fig:m11}
\end{figure*}

The second example, featuring $\mwd=1.1\,\msol$ with $ \zig=0.5 $ ($\Delta x_{\rm{min}}=1\,\rm{km}$ and $\Rig=100\,\rm{km}$), as presented in Fig.~\ref{fig:m11}, appears notably simpler. Owing to the higher density around the ignition zone, \nick is consistently produced behind the TNDW front immediately after ignition, as evident in panel (a). Consequently, the \nick front appears almost spherical, except for the most northern region, where the TNDW encounters lower densities. Panel (b) shows the structure around the breakout time of the northern side of the WD. In contrast to the $\mwd=0.8\,\msol$ case, regions closer to the ignition location are breached first. The segmentation of the ejecta, mentioned above, is more evident in panel (b) compared to panel (c) of Fig.~\ref{fig:m08}.

Panel (c) illustrates the ejecta at the homologous expansion stage. Unlike the $\mwd=0.8\,\msol$ case, no significant region in the ejecta is associated with the early ignition stage. The ejecta appears more spherical, and the \nick mass is distributed in a single region. This example results in $\mnii=0.78\,\msol$ and $t_0=30.0$ days, compared to $\mnii=0.80\,\msol$ and $t_0=30.3$ d for the 2D symmetric case. The ignition location has a much lower impact on the results due to the higher mass of the WD. This is attributed to the stability and speed of the TNDW, primarily determined by the initial densities, which are higher in more massive WDs.

\subsection{Convergence study}
\label{sec:exconv}

\begin{figure}
    \includegraphics[width=\columnwidth]{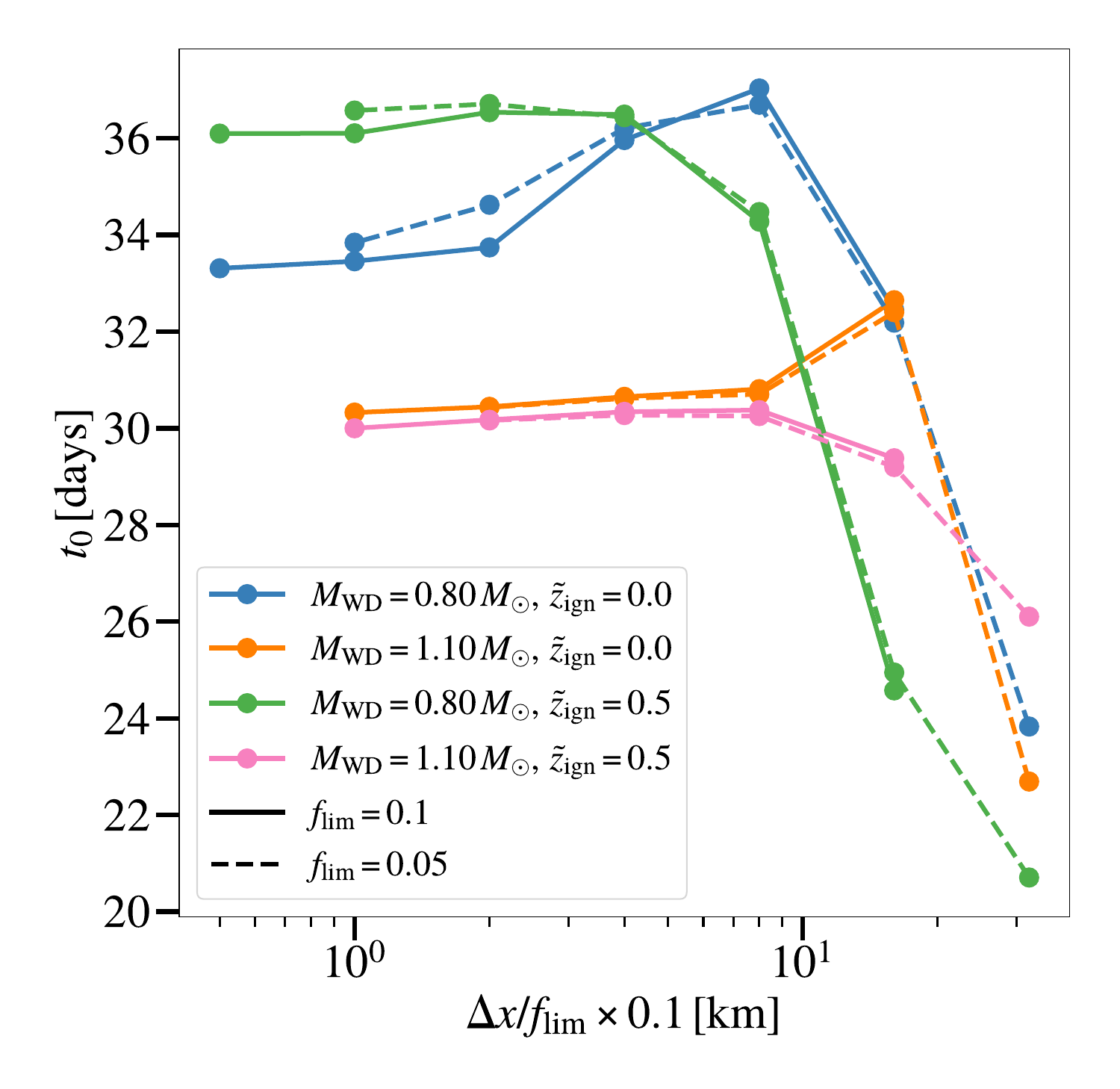}{(a)}
    \includegraphics[width=\columnwidth]{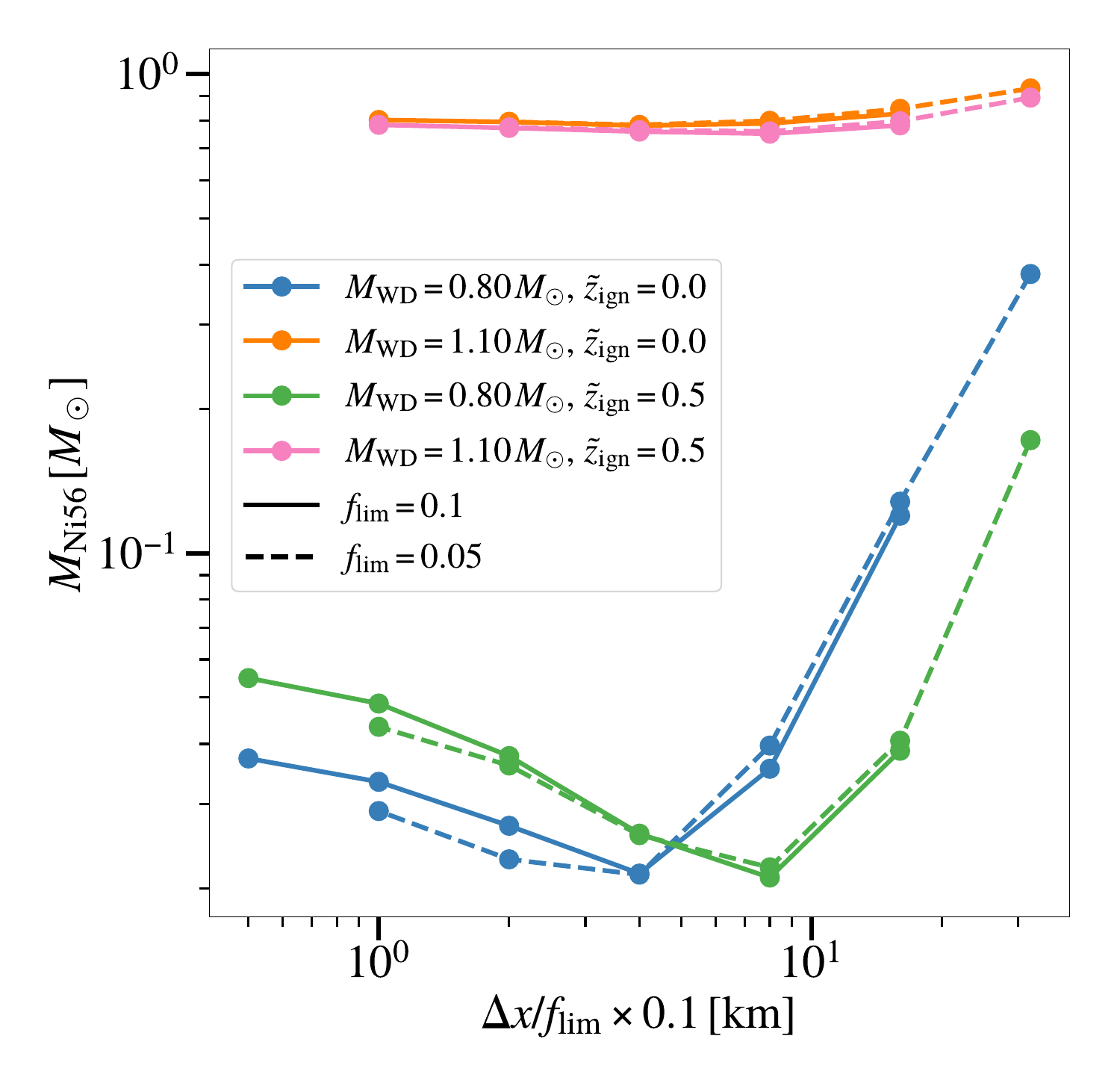}{(b)}
    \caption{The convergence of $t_0$ (panel a) and \mni (panel b) as a function of the normalized resolution $\Delta x_{\rm{min}}/f_{\rm{lim}}\times0.1$ for $\mwd=0.8\,\msol$ ($\zig=0$ in blue and $\zig=0.5$ in green) and for $\mwd=1.1\,\msol$ ($\zig=0$ in orange and $\zig=0.5$ in pink). The solid (dashed) lines represent the results for $\flim=0.1$ ($\flim=0.05$). As expected, the scaled results are roughly independent of $\flim$. A resolution of $\Delta x_{\rm{min}}(\flim/0.1)=1\,\rm{km}$ is sufficient for $t_0$ convergence to the percentage level. At this resolution, $\mnii$ for the $\mwd=1.1\,\msol$ case converges to the percentage level, while for the $\mwd=0.8\,\msol$ case, it converges to about $\mysim 20\%$. For the $\mwd=0.8\,\msol$ case, $\mnii$ converges to about $\mysim 10\%$ with a resolution of $\Delta x_{\rm{min}}=0.5\,\rm{km}$.}
    \label{fig:flim}.
\end{figure}

Fig.~\ref{fig:flim} illustrates the convergence of $t_0$ and \mni for the example cases discussed in Sections~\ref{sec:ex08} and~\ref{sec:ex11}, along with similar cases of 2D central ignition, for two burning limiter values: the default of this work, which is $\flim=0.1$, and $\flim=0.05$. Following the approach taken by \refKWSs, we show the results as a function of the normalized resolution with respect to the burning limiter, as the burning limiter utilizes approximately $\mysim 1/\flim $ cells to describe the fast-burning region. Indeed, the scaled results are roughly independent of $\flim$. We find that a resolution of $\Delta x_{\rm{min}}(\flim/0.1)=1\,\rm{km}$ is adequate for convergence of $t_0$ to the percentage level. For the same resolution, the $\mnii$ of the $\mwd=1.1\,\msol$ case is converged to the percentage level, but the $\mnii$ of the $\mwd=0.8\,\msol$ case only converges to $\mysim 20\%$. The $\mnii$ of the $\mwd=0.8\,\msol$ case converges to $\mysim 10\%$ with a resolution of $\Delta x_{\rm{min}}=0.5\,\rm{km}$. \refKWS obtained a similar result for their 1D simulation, supporting our estimates for the convergence level of our results. \refKWS found that a resolution  of $\mysim 0.1\,\rm{km}$ was necessary to achieve percent convergence in \mni for low-mass WDs. Achieving such high resolution in 2D simulation is challenging and unnecessary for this work. We, therefore, use $\Delta x_{\min}=1\,\rm{km}$ for high-mass, $\mwd\ge 0.95$, WDs and $\Delta x_{\min}=0.5\,\rm{km}$ for low-mass, $\mwd\le 0.9$, WDs throughout this work, allowing a convergence of $t_0$ to the percentage level and of $\mnii$ to $\mysim 10\%$.


We recalculate the cases presented in Fig.~\ref{fig:flim} with $\Delta x_{\rm{min}}=1\,\rm{km}$ and the 69-isotopes list calibrated by \refKWSs, for which the SCD $t_0-$\mni relation is accurately calculated. The impact of employing a more detailed isotope network on the outcomes is not substantial (see also Appendix~\ref{sec:oned}). In the more sensitive case involving off-centre ignition of low-mass WD, the effect on \mni is $\lesssim25\,\%$ while it is lower for the other cases ($\mysim10\,\%$ for the symmetric $\mwd=0.8\,\msol$ case and $\mysim1\,\%$ for the $\mwd=1.1\,\msol$ cases). The effect on $t_0$ is even smaller ($<2\,\%$), suggesting that our default 38-isotope list is adequate for comparing the DDM calculations with the observed $ t_0-\mnii $ relation. The impact of other numerical parameters is relatively small (see Appendix~\ref{sec:numericsens}).


\section{Results}
\label{sec:results}

\begin{figure}
\includegraphics[width=\columnwidth]{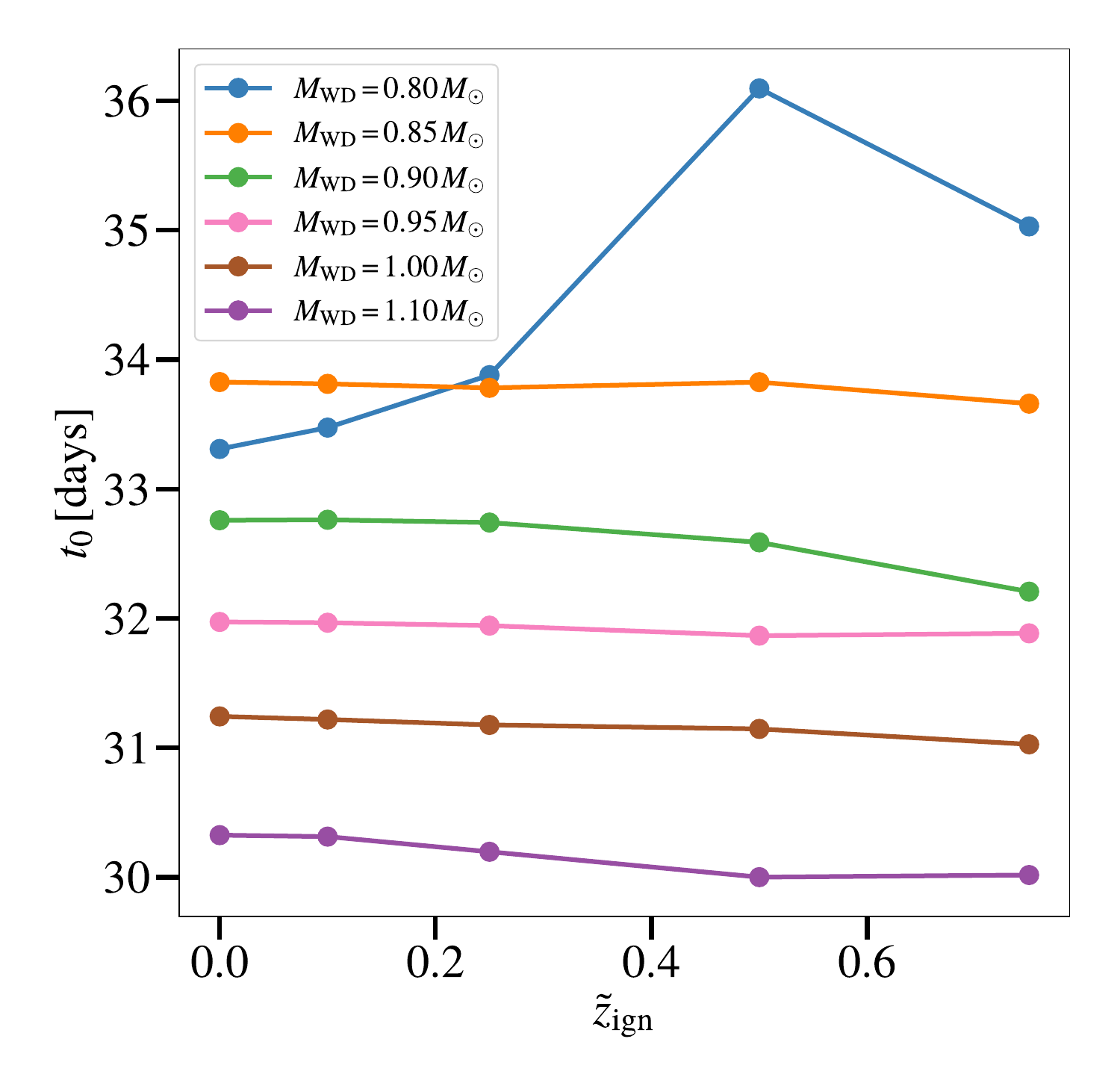}{(a)}
\includegraphics[width=\columnwidth]{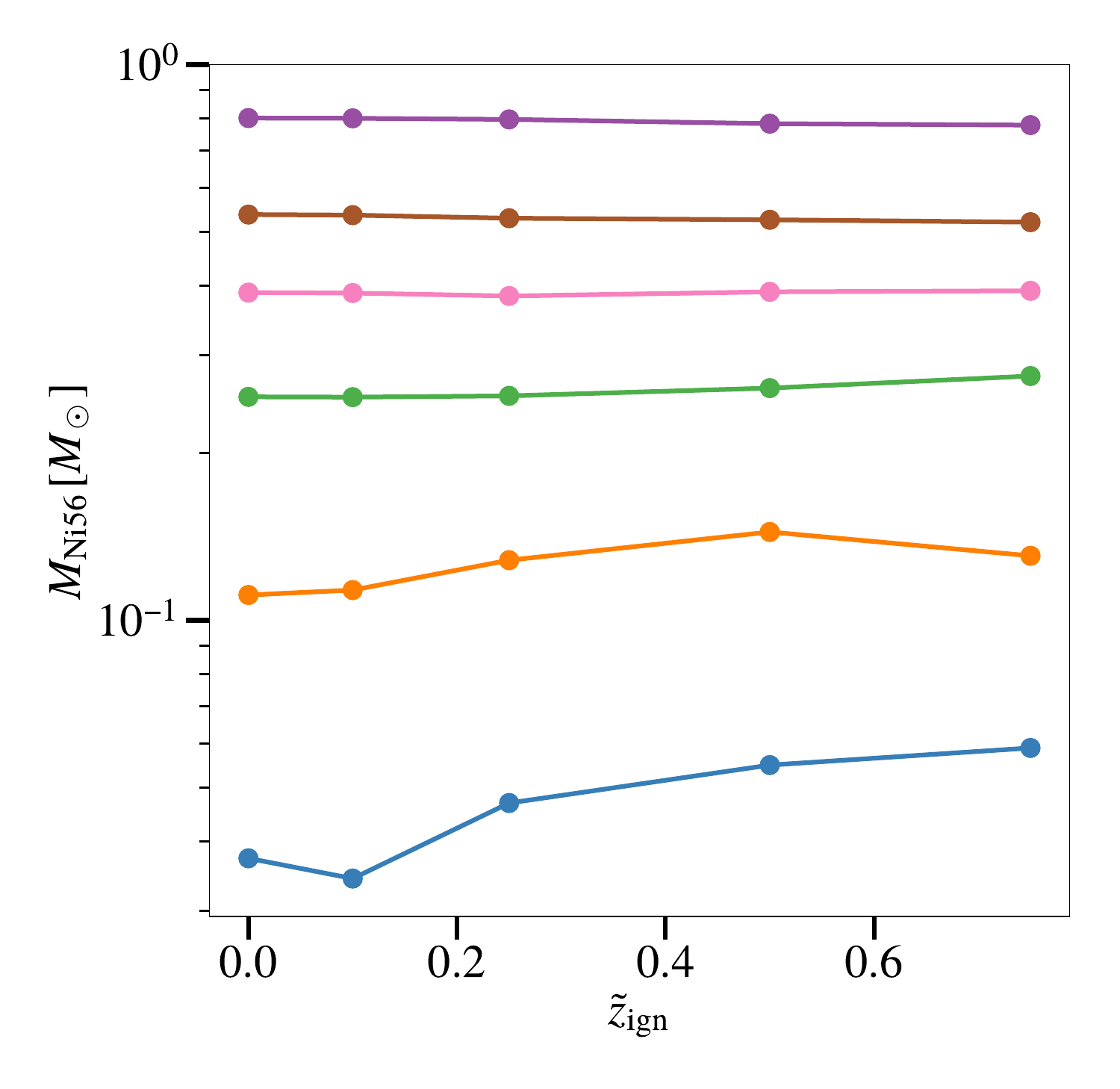}{(b)}
\caption{The obtained values of $t_0$ (a) and \mni (b), as a function of the ignition location from the simulations with the highest resolution for each parameter set. The different colours (blue, orange, green, pink, brown, and purple) represent different $\mwd$ ($\mwd=0.8,0.85,0.9,0.95,1.0,1.1\msol$, respectively). Both $t_0$ and \mni are primarily influenced by $\mwd$ and exhibit only a slight dependence on the ignition location. Exceptions are light WDs, which exhibit a $10\%(5\%)$ variance in $t_0$ and $80\%(45\%)$ variance in \mni for $\mwd=0.85(0.8)\,M_{\odot}$, linked to the geometry of the TNDW as it traverses the central regions of the WD (see text).}
\label{fig:results}
\end{figure}

In this section, we discuss the results of our simulations for different values of $\mwd$ and $\zig$. The results of $t_0$ and \mni for each simulation in this work are given in Table~\ref{tab:tabres}. Fig.~\ref{fig:results} shows the results from the simulations with the highest resolution for each parameter set. Both $t_0$ and \mni are primarily influenced by $\mwd$ and exhibit only a slight dependence on the ignition location. The only exceptions are light WDs with $10\%(1\%)$ variance in $t_0$ and $60\%(25\%)$ variance in \mni for $\mwd=0.8(0.85)\,M_{\odot}$. Note that in both cases, only a small amount of \nick is synthesized (compared to the majority of SNe Ia).

The observed increase in \mni with rising $\zig$ for light WDs is linked to the geometry of the TNDW as it traverses the central regions of the WD. In the symmetric scenario, where the TNDW is ignited at the centre, it propagates outward as a diverging spherical wave. Conversely, in the off-centre ignition scenario, the TNDW moves toward the centre from an external ignition point, resembling a planar TNDW. Only a small fraction of the central mass is converted into $^{56}$Ni for light WDs. The difference in TNDW geometry affects the TNDW velocity and downstream temperature, influencing the efficiency of $^{56}$Ni burning. The planar TNDW is stronger than the diverging spherical TNDW, leading to a more complete $^{56}$Ni synthesis in the planar case. This explains the observed increase in \mni with rising $\zig$. Consequently, $t_0$ also increases [see panel (a)] as a larger fraction of $^{56}$Ni is concentrated in the central parts of the ejecta.

Panels (a) and (b) of Fig.~\ref{fig:results} underscore the primary finding of this study, as depicted in Fig.~\ref{fig:main_plot}. Our findings suggest a minor dependence on the ignition location, echoing the 1D computations of \refKWSs. As a result, the disparity with the observed $t_0$-\mni relation persists without resolution.


\section{Summary and Discussion}
\label{sec:summary}

Utilizing a modified version of \textsc{FLASH4.0} with an accurate and efficient burning scheme, we conducted simulations of asymmetrical WD detonations resulting from off-centre ignition. Our findings indicate that \mni and $t_0$ exhibit only slight dependencies on the ignition location. Consequently, the $t_0$-\mni relationship we observe closely resembles the 1D results obtained by \refKWSs. This implies that incorporating this specific 2D aspect of the DDM model does not alleviate the tension with observations identified for the 1D SCD model. Most of our results are shown to be converged to the level of $\mysim1\%$, except when $\mnii$ is relatively small ($<0.1\,\msol$) for low-mass WDs.

We have not accounted for the compression wave that propagates in the CO core prior to ignition, which alters the initial conditions of the CO core and has relied on the assumption of hydrostatic equilibrium, which is a simplification. To comprehensively address this effect, more sophisticated simulations involving the detonation of the He shell are necessary. Such simulations were carried out by \citet[][]{boos_multidimensional_2021}, and \citet[][]{boos_type_2024} derived the $t_0$-\mni relation from these simulations, shown as red points in Fig.~\ref{fig:main_plot}. Their results align closely with those obtained using our simplified model, suggesting that the compression wave in the CO core has a minimal effect on the final results. However, \citet[][]{boos_multidimensional_2021} did not demonstrate that their simulation resolution, $\Delta x_{\rm{min}}=4\,\rm{km}$, is sufficient for convergence, and they deactivated the burning limiter during ignition in the CO core. Future work should address these limitations by calculating this configuration with higher numerical resolution (at least $\Delta x_{\rm{min}}=1\,\rm{km}$, as we have shown) and resolving the ignition in the CO core. Additionally, the outcomes may be influenced by the interaction of the ejecta with the WD's companion, which inherently possesses a 3D nature. This interaction imposes further computational demands, as highlighted by \citet[][]{boos_type_2024}.

The $t_0$ values presented in this study represent a 1D average over various viewing angles, which can be observationally derived only at late times ($\gtrsim$1 yr). We compared our calculations to values determined from observations at $\mysim100\,\textrm{d}$, where viewing-angle dependence is anticipated. Given the minimal asymmetries in the ejecta in our study and the small deviations of the average $t_0$ values from centrally-ignited values, relying on the average $t_0$ approximation appears reasonable. We defer a more exhaustive exploration of the viewing-angle effect to future investigations.

\section*{Acknowledgements}

We thank Amir Sharon for performing the Monte-Carlo $\gamma$-ray transport calculations and Boaz Katz for the useful discussions. Doron Kushnir is supported by a research grant from The Abramson Family Center for Young Scientists, and by the Minerva Stiftung.

\section*{Data availability}

All the data of the ejecta obtained from the calculation presented in this paper is publicly available in \url{https://drive.google.com/drive/folders/1RzQzcvCYmVnN_Eq8oiZtNd8kJ9kMhvyQ?usp=sharing}.

\bibliography{references,extra_references}{}

\begin{thebibliography}{}
\makeatletter
\relax
\def\mn@urlcharsother{\let\do\@makeother \do\$\do\&\do\#\do\^\do\_\do\%\do\~}
\def\mn@doi{\begingroup\mn@urlcharsother \@ifnextchar [ {\mn@doi@}
  {\mn@doi@[]}}
\def\mn@doi@[#1]#2{\def\@tempa{#1}\ifx\@tempa\@empty \href
  {http://dx.doi.org/#2} {doi:#2}\else \href {http://dx.doi.org/#2} {#1}\fi
  \endgroup}
\def\mn@eprint#1#2{\mn@eprint@#1:#2::\@nil}
\def\mn@eprint@arXiv#1{\href {http://arxiv.org/abs/#1} {{\tt arXiv:#1}}}
\def\mn@eprint@dblp#1{\href {http://dblp.uni-trier.de/rec/bibtex/#1.xml}
  {dblp:#1}}
\def\mn@eprint@#1:#2:#3:#4\@nil{\def\@tempa {#1}\def\@tempb {#2}\def\@tempc
  {#3}\ifx \@tempc \@empty \let \@tempc \@tempb \let \@tempb \@tempa \fi \ifx
  \@tempb \@empty \def\@tempb {arXiv}\fi \@ifundefined
  {mn@eprint@\@tempb}{\@tempb:\@tempc}{\expandafter \expandafter \csname
  mn@eprint@\@tempb\endcsname \expandafter{\@tempc}}}

\bibitem[\protect\citeauthoryear{Bildsten, Shen, Weinberg  \&
  Nelemans}{Bildsten et~al.}{2007}]{bildsten_faint_2007}
Bildsten L.,  Shen K.~J.,  Weinberg N.~N.,   Nelemans G.,  2007, \mn@doi [The
  Astrophysical Journal] {10.1086/519489}, 662, L95

\bibitem[\protect\citeauthoryear{Boos, Townsley, Shen, Caldwell  \& Miles}{Boos
  et~al.}{2021}]{boos_multidimensional_2021}
Boos S.~J.,  Townsley D.~M.,  Shen K.~J.,  Caldwell S.,   Miles B.~J.,  2021,
  \mn@doi [The Astrophysical Journal] {10.3847/1538-4357/ac07a2}, 919, 126

\bibitem[\protect\citeauthoryear{Boos, Townsley  \& Shen}{Boos
  et~al.}{2024}]{boos_type_2024}
Boos S.~J.,  Townsley D.~M.,   Shen K.~J.,  2024, \mn@doi [The Astrophysical
  Journal] {10.3847/1538-4357/ad5da2}, 972, 200

\bibitem[\protect\citeauthoryear{Burmester, Ferrario, Pakmor, Seitenzahl,
  Ruiter  \& Hole}{Burmester et~al.}{2023}]{burmester_arepo_2023}
Burmester U.~P.,  Ferrario L.,  Pakmor R.,  Seitenzahl I.~R.,  Ruiter A.~J.,
  Hole M.,  2023, \mn@doi [Monthly Notices of the Royal Astronomical Society]
  {10.1093/mnras/stad1394}, p. stad1394

\bibitem[\protect\citeauthoryear{{Chen} et~al.,}{{Chen}
  et~al.}{2023}]{Chen2023}
{Chen} N.~M.,  et~al., 2023, \mn@doi [\apjl] {10.3847/2041-8213/acb6d8}, \href
  {https://ui.adsabs.harvard.edu/abs/2023ApJ...944L..28C} {944, L28}

\bibitem[\protect\citeauthoryear{Dubey, Reid, Weide, Antypas, Ganapathy, Riley,
  Sheeler  \& Siegal}{Dubey et~al.}{2009}]{dubey_extensible_2009}
Dubey A.,  Reid L.~B.,  Weide K.,  Antypas K.,  Ganapathy M.~K.,  Riley K.,
  Sheeler D.,   Siegal A.,  2009, \mn@doi [Parallel Computing]
  {10.1016/j.parco.2009.08.001}, 35, 512

\bibitem[\protect\citeauthoryear{Fink, Hillebrandt  \& Röpke}{Fink
  et~al.}{2007}]{fink_double-detonation_2007}
Fink M.,  Hillebrandt W.,   Röpke F.~K.,  2007, \mn@doi [Astronomy and
  Astrophysics] {10.1051/0004-6361:20078438}, 476, 1133

\bibitem[\protect\citeauthoryear{Fink, Röpke, Hillebrandt, Seitenzahl, Sim  \&
  Kromer}{Fink et~al.}{2010}]{fink_double-detonation_2010}
Fink M.,  Röpke F.~K.,  Hillebrandt W.,  Seitenzahl I.~R.,  Sim S.~A.,
  Kromer M.,  2010, \mn@doi [Astronomy and Astrophysics]
  {10.1051/0004-6361/200913892}, 514, A53

\bibitem[\protect\citeauthoryear{{Fremling} et~al.,}{{Fremling}
  et~al.}{2020}]{Fremling2020}
{Fremling} C.,  et~al., 2020, \mn@doi [\apj] {10.3847/1538-4357/ab8943}, \href
  {https://ui.adsabs.harvard.edu/abs/2020ApJ...895...32F} {895, 32}

\bibitem[\protect\citeauthoryear{Fryxell et~al.,}{Fryxell
  et~al.}{2000}]{fryxell_flash_2000}
Fryxell B.,  et~al., 2000, \mn@doi [The Astrophysical Journal Supplement
  Series] {10.1086/317361}, 131, 273

\bibitem[\protect\citeauthoryear{Ghosh \& Kushnir}{Ghosh \&
  Kushnir}{2022}]{ghosh_confronting_2022}
Ghosh A.,  Kushnir D.,  2022, \mn@doi [Monthly Notices of the Royal
  Astronomical Society] {10.1093/mnras/stac1846}, 515, 286

\bibitem[\protect\citeauthoryear{{Glasner}, {Livne}, {Steinberg}, {Yalinewich}
  \& {Truran}}{{Glasner} et~al.}{2018}]{Glasner2018}
{Glasner} S.~A.,  {Livne} E.,  {Steinberg} E.,  {Yalinewich} A.,   {Truran}
  J.~W.,  2018, \mn@doi [\mnras] {10.1093/mnras/sty421}, \href
  {https://ui.adsabs.harvard.edu/abs/2018MNRAS.476.2238G} {476, 2238}

\bibitem[\protect\citeauthoryear{Gronow, Collins, Ohlmann, Pakmor, Kromer,
  Seitenzahl, Sim  \& Röpke}{Gronow et~al.}{2020}]{gronow_sne_2020}
Gronow S.,  Collins C.,  Ohlmann S.~T.,  Pakmor R.,  Kromer M.,  Seitenzahl
  I.~R.,  Sim S.~A.,   Röpke F.~K.,  2020, \mn@doi [Astronomy and
  Astrophysics] {10.1051/0004-6361/201936494}, 635, A169

\bibitem[\protect\citeauthoryear{Gronow, Collins, Sim  \& Röpke}{Gronow
  et~al.}{2021a}]{gronow_double_2021}
Gronow S.,  Collins C.~E.,  Sim S.~A.,   Röpke F.~K.,  2021a, \mn@doi
  [Astronomy \& Astrophysics] {10.1051/0004-6361/202039954}, 649, A155

\bibitem[\protect\citeauthoryear{Gronow, Côté, Lach, Seitenzahl, Collins, Sim
   \& Röpke}{Gronow et~al.}{2021b}]{gronow_metallicity-dependent_2021}
Gronow S.,  Côté B.,  Lach F.,  Seitenzahl I.~R.,  Collins C.~E.,  Sim S.~A.,
    Röpke F.~K.,  2021b, \mn@doi [Astronomy \& Astrophysics]
  {10.1051/0004-6361/202140881}, 656, A94

\bibitem[\protect\citeauthoryear{Hoeflich \& Khokhlov}{Hoeflich \&
  Khokhlov}{1996}]{hoeflich_explosion_1996}
Hoeflich P.,  Khokhlov A.,  1996, \mn@doi [The Astrophysical Journal]
  {10.1086/176748}, 457, 500

\bibitem[\protect\citeauthoryear{{Jacobs}, {Zingale}, {Nonaka}, {Almgren}  \&
  {Bell}}{{Jacobs} et~al.}{2016}]{Jacobs2016}
{Jacobs} A.~M.,  {Zingale} M.,  {Nonaka} A.,  {Almgren} A.~S.,   {Bell} J.~B.,
  2016, \mn@doi [\apj] {10.3847/0004-637X/827/1/84}, \href
  {https://ui.adsabs.harvard.edu/abs/2016ApJ...827...84J} {827, 84}

\bibitem[\protect\citeauthoryear{Jeffery}{Jeffery}{1999}]{jeffery_radioactive_1999}
Jeffery D.~J.,  1999, \mn@doi [arXiv e-prints]
  {10.48550/arXiv.astro-ph/9907015}, pp astro--ph/9907015

\bibitem[\protect\citeauthoryear{Katz, Kushnir  \& Dong}{Katz
  et~al.}{2013}]{katz_exact_2013}
Katz B.,  Kushnir D.,   Dong S.,  2013, Technical report, An exact integral
  relation between the {Ni56} mass and the bolometric light curve of a type
  {Ia} supernova, \url {https://ui.adsabs.harvard.edu/abs/2013arXiv1301.6766K},
  \mn@doi{10.48550/arXiv.1301.6766.
}, \url {https://ui.adsabs.harvard.edu/abs/2013arXiv1301.6766K}

\bibitem[\protect\citeauthoryear{Khokhlov}{Khokhlov}{1993}]{khokhlov_stability_1993}
Khokhlov A.~M.,  1993, \mn@doi [The Astrophysical Journal] {10.1086/173475},
  419, 200

\bibitem[\protect\citeauthoryear{Kromer, Sim, Fink, Röpke, Seitenzahl  \&
  Hillebrandt}{Kromer et~al.}{2010}]{kromer_double-detonation_2010}
Kromer M.,  Sim S.~A.,  Fink M.,  Röpke F.~K.,  Seitenzahl I.~R.,
  Hillebrandt W.,  2010, \mn@doi [The Astrophysical Journal]
  {10.1088/0004-637X/719/2/1067}, 719, 1067

\bibitem[\protect\citeauthoryear{Kushnir \& Katz}{Kushnir \&
  Katz}{2020}]{kushnir_accurate_2020}
Kushnir D.,  Katz B.,  2020, \mn@doi [Monthly Notices of the Royal Astronomical
  Society] {10.1093/mnras/staa594}, 493, 5413

\bibitem[\protect\citeauthoryear{Kushnir, Wygoda  \& Sharon}{Kushnir
  et~al.}{2020}]{kushnir_sub-chandrasekhar-mass_2020}
Kushnir D.,  Wygoda N.,   Sharon A.,  2020, \mn@doi [Monthly Notices of the
  Royal Astronomical Society] {10.1093/mnras/staa3017}, 499, 4725

\bibitem[\protect\citeauthoryear{Livne}{Livne}{1990}]{livne_successive_1990}
Livne E.,  1990, \mn@doi [The Astrophysical Journal] {10.1086/185721}, 354, L53

\bibitem[\protect\citeauthoryear{Maoz, Mannucci  \& Nelemans}{Maoz
  et~al.}{2014}]{maoz_observational_2014}
Maoz D.,  Mannucci F.,   Nelemans G.,  2014, \mn@doi [Annual Review of
  Astronomy and Astrophysics] {10.1146/annurev-astro-082812-141031}, 52, 107

\bibitem[\protect\citeauthoryear{Moore, Townsley  \& Bildsten}{Moore
  et~al.}{2013}]{moore_effects_2013}
Moore K.,  Townsley D.~M.,   Bildsten L.,  2013, \mn@doi [The Astrophysical
  Journal] {10.1088/0004-637X/776/2/97}, 776, 97

\bibitem[\protect\citeauthoryear{Nomoto}{Nomoto}{1982a}]{nomoto_accreting_1982-1}
Nomoto K.,  1982a, \mn@doi [The Astrophysical Journal] {10.1086/159682}, 253,
  798

\bibitem[\protect\citeauthoryear{Nomoto}{Nomoto}{1982b}]{nomoto_accreting_1982}
Nomoto K.,  1982b, \mn@doi [The Astrophysical Journal] {10.1086/160031}, 257,
  780

\bibitem[\protect\citeauthoryear{Nugent, Baron, Branch, Fisher  \&
  Hauschildt}{Nugent et~al.}{1997}]{nugent_synthetic_1997}
Nugent P.,  Baron E.,  Branch D.,  Fisher A.,   Hauschildt P.~H.,  1997,
  \mn@doi [The Astrophysical Journal] {10.1086/304459}, 485, 812

\bibitem[\protect\citeauthoryear{{Papatheodore} \& {Messer}}{{Papatheodore} \&
  {Messer}}{2014}]{Papatheodore2014}
{Papatheodore} T.~L.,  {Messer} O.~E.~B.,  2014, \mn@doi [\apj]
  {10.1088/0004-637X/782/1/12}, \href
  {https://ui.adsabs.harvard.edu/abs/2014ApJ...782...12P} {782, 12}

\bibitem[\protect\citeauthoryear{{Parete-Koon}, {Smith}, {Guidry}, {Hix}  \&
  {Messer}}{{Parete-Koon} et~al.}{2012}]{Parete2012}
{Parete-Koon} S.~T.,  {Smith} C.~R.,  {Guidry} M.~W.,  {Hix} W.~R.,   {Messer}
  O.~E.~B.,  2012, in Journal of Physics Conference Series. IOP, p. 012028,
  \mn@doi{10.1088/1742-6596/402/1/012028}

\bibitem[\protect\citeauthoryear{Paxton, Bildsten, Dotter, Herwig, Lesaffre  \&
  Timmes}{Paxton et~al.}{2011}]{2011ApJS..192....3P}
Paxton B.,  Bildsten L.,  Dotter A.,  Herwig F.,  Lesaffre P.,   Timmes F.,
  2011, \mn@doi [The Astrophysical Journal Supplement Series]
  {10.1088/0067-0049/192/1/3}, 192, 3

\bibitem[\protect\citeauthoryear{{Perley} et~al.,}{{Perley}
  et~al.}{2020}]{Perley2020}
{Perley} D.~A.,  et~al., 2020, \mn@doi [\apj] {10.3847/1538-4357/abbd98}, \href
  {https://ui.adsabs.harvard.edu/abs/2020ApJ...904...35P} {904, 35}

\bibitem[\protect\citeauthoryear{Phillips}{Phillips}{1993}]{phillips_absolute_1993}
Phillips M.~M.,  1993, \mn@doi [The Astrophysical Journal] {10.1086/186970},
  413, L105

\bibitem[\protect\citeauthoryear{Piersanti, Yungelson  \& Bravo}{Piersanti
  et~al.}{2024}]{piersanti_expected_2024}
Piersanti L.,  Yungelson L.~R.,   Bravo E.,  2024, \mn@doi [Astronomy \&
  Astrophysics] {10.1051/0004-6361/202450008}, 689, A287

\bibitem[\protect\citeauthoryear{Polin, Nugent  \& Kasen}{Polin
  et~al.}{2019}]{polin_observational_2019}
Polin A.,  Nugent P.,   Kasen D.,  2019, \mn@doi [The Astrophysical Journal]
  {10.3847/1538-4357/aafb6a}, 873, 84

\bibitem[\protect\citeauthoryear{{Scalzo} et~al.,}{{Scalzo}
  et~al.}{2014}]{Scalzo2014}
{Scalzo} R.,  et~al., 2014, \mn@doi [\mnras] {10.1093/mnras/stu350}, \href
  {https://ui.adsabs.harvard.edu/abs/2014MNRAS.440.1498S} {440, 1498}

\bibitem[\protect\citeauthoryear{Sharon}{Sharon}{2023}]{sharon_personal_2023}
Sharon A.,  2023, Personal communication

\bibitem[\protect\citeauthoryear{Sharon \& Kushnir}{Sharon \&
  Kushnir}{2020}]{sharon_-ray_2020}
Sharon A.,  Kushnir D.,  2020, \mn@doi [Monthly Notices of the Royal
  Astronomical Society] {10.1093/mnras/staa1745}, 496, 4517

\bibitem[\protect\citeauthoryear{{Sharon} \& {Kushnir}}{{Sharon} \&
  {Kushnir}}{2022}]{SK2021}
{Sharon} A.,  {Kushnir} D.,  2022, \mn@doi [\mnras] {10.1093/mnras/stab3380},
  \href {https://ui.adsabs.harvard.edu/abs/2022MNRAS.509.5275S} {509, 5275}

\bibitem[\protect\citeauthoryear{Sharon \& Kushnir}{Sharon \&
  Kushnir}{2024}]{sharon_all_2024}
Sharon A.,  Kushnir D.,  2024, All known {Type} {Ia} supernovae models fail to
  reproduce the observed bolometric luminosity-width correlation,
  \mn@doi{10.48550/arXiv.2407.06859}, \url {http://arxiv.org/abs/2407.06859}

\bibitem[\protect\citeauthoryear{Shen \& Bildsten}{Shen \&
  Bildsten}{2014}]{shen_ignition_2014}
Shen K.~J.,  Bildsten L.,  2014, \mn@doi [The Astrophysical Journal]
  {10.1088/0004-637X/785/1/61}, 785, 61

\bibitem[\protect\citeauthoryear{Shen \& Moore}{Shen \&
  Moore}{2014}]{shen_initiation_2014}
Shen K.~J.,  Moore K.,  2014, \mn@doi [The Astrophysical Journal]
  {10.1088/0004-637X/797/1/46}, 797, 46

\bibitem[\protect\citeauthoryear{{Shen}, {Toonen}  \& {Graur}}{{Shen}
  et~al.}{2017}]{Shen2017}
{Shen} K.~J.,  {Toonen} S.,   {Graur} O.,  2017, \mn@doi [\apjl]
  {10.3847/2041-8213/aaa015}, \href
  {https://ui.adsabs.harvard.edu/abs/2017ApJ...851L..50S} {851, L50}

\bibitem[\protect\citeauthoryear{Shen, Boos  \& Townsley}{Shen
  et~al.}{2024}]{shen_almost_2024}
Shen K.~J.,  Boos S.~J.,   Townsley D.~M.,  2024, \mn@doi [The Astrophysical
  Journal] {10.3847/1538-4357/ad7379}, 975, 127

\bibitem[\protect\citeauthoryear{{Stritzinger}, {Leibundgut}, {Walch}  \&
  {Contardo}}{{Stritzinger} et~al.}{2006}]{Stritzinger2006}
{Stritzinger} M.,  {Leibundgut} B.,  {Walch} S.,   {Contardo} G.,  2006,
  \mn@doi [\aap] {10.1051/0004-6361:20053652}, \href
  {https://ui.adsabs.harvard.edu/abs/2006A&A...450..241S} {450, 241}

\bibitem[\protect\citeauthoryear{Swartz, Sutherland  \& Harkness}{Swartz
  et~al.}{1995}]{swartz_gamma-ray_1995}
Swartz D.~A.,  Sutherland P.~G.,   Harkness R.~P.,  1995, \mn@doi [The
  Astrophysical Journal] {10.1086/175834}, 446, 766

\bibitem[\protect\citeauthoryear{Timmes et~al.,}{Timmes
  et~al.}{2000}]{timmes_cellular_2000}
Timmes F.~X.,  et~al., 2000, \mn@doi [The Astrophysical Journal]
  {10.1086/317135}, 543, 938

\bibitem[\protect\citeauthoryear{Townsley, Miles, Shen  \& Kasen}{Townsley
  et~al.}{2019}]{townsley_double_2019}
Townsley D.~M.,  Miles B.~J.,  Shen K.~J.,   Kasen D.,  2019, \mn@doi [The
  Astrophysical Journal] {10.3847/2041-8213/ab27cd}, 878, L38

\bibitem[\protect\citeauthoryear{Woosley \& Kasen}{Woosley \&
  Kasen}{2011}]{woosley_sub-chandrasekhar_2011}
Woosley S.~E.,  Kasen D.,  2011, \mn@doi [The Astrophysical Journal]
  {10.1088/0004-637X/734/1/38}, 734, 38

\bibitem[\protect\citeauthoryear{Woosley \& Weaver}{Woosley \&
  Weaver}{1994}]{woosley_sub--chandrasekhar_1994}
Woosley S.~E.,  Weaver T.~A.,  1994, \mn@doi [The Astrophysical Journal]
  {10.1086/173813}, 423, 371

\bibitem[\protect\citeauthoryear{Wygoda, Elbaz  \& Katz}{Wygoda
  et~al.}{2019}]{wygoda_type_2019}
Wygoda N.,  Elbaz Y.,   Katz B.,  2019, \mn@doi [Monthly Notices of the Royal
  Astronomical Society] {10.1093/mnras/stz145}, 484, 3941

\bibitem[\protect\citeauthoryear{{Zingale}, {Nonaka}, {Almgren}, {Bell},
  {Malone}  \& {Orvedahl}}{{Zingale} et~al.}{2013}]{Zingale2013}
{Zingale} M.,  {Nonaka} A.,  {Almgren} A.~S.,  {Bell} J.~B.,  {Malone} C.~M.,
  {Orvedahl} R.~J.,  2013, \mn@doi [\apj] {10.1088/0004-637X/764/1/97}, \href
  {https://ui.adsabs.harvard.edu/abs/2013ApJ...764...97Z} {764, 97}

\makeatother
\end{thebibliography}
\bibliographystyle{mnras}

\bsp	

\appendix


\section{AMR scheme}
\label{sec:amr}

This appendix describes the AMR scheme we implemented in our simulations. As outlined in Section~\ref{sec:modif}, we divide the computational grid into 50 discrete angles centred around the ignition point. For each angle and time step, we determine the outgoing shock radius, denoted as $r_{\rm{shock}}$, as the largest distance (relative to the ignition point) where the temperature exceeds the initial temperature by 1\%. Additionally, at each angle, we determine the breakout time, $t_{\rm{breakout}}$, as the moment when $r_{\rm{shock}}$ first exceeds $r_{\rm{WD}}-20\Delta x_{\rm{min}}$. Here, $r_{\rm{WD}}$ represents the maximum distance from the ignition point to the WD's surface within a specific angle. The maximal breakout time over all angles is $t_{\rm{breakout,max}}$.

During each simulation step, we compute the following parameters for every computational block:
\begin{enumerate}[wide, labelwidth=!,itemindent=!,labelindent=0pt, leftmargin=0em, parsep=0pt]
    \item The minimal radius, relative to the ignition location, calculated at the centre of the cells, denoted as $r_{\rm{min}}$.
    \item The maximal density, represented by $\rho_{\rm{max}}$.
    \item The minimal ratio of the typical burning time-scale to the sound crossing time of the cells, denoted as $g_{\rm{fac}}$.
    \item $ g_{\rm{fac,2}}$ which is $g_{\rm{fac}}$ for coarser resolution by a factor of two.
\end{enumerate}
These values are used in our AMR scheme, which operates based on the following criteria (each condition supersedes all previous conditions):
    \begin{enumerate}[wide, labelwidth=!,itemindent=!,labelindent=0pt, leftmargin=0em, parsep=0pt]
\item A density gradient refinement condition with \texttt{refine\_cutoff} = 0.8, \texttt{derefine\_cutoff} = 0.2, and
\texttt{refine\_filter} = 0.01.
\item If $t>t_{\rm{breakout,max}}$, $\rho_{\rm{max}}<10^5\,\rm{g/cm^3}$ and the refinement level is greater than \texttt{lrefine\_floor}, derefine.
\item If $t<t_{\rm{breakout,max}}$, $\rho_{\rm{max}}<10^5\,\rm{g/cm^3}$, the refinement level is greater than \texttt{lrefine\_floor}, and $r_{\rm{min}}>r_{\rm{shock}}$, derefine.
\item If $t<t_{\rm{breakout,max}}$, $r_{\rm{min}}>r_{\rm{WD}}$, and the refinement level is greater than \texttt{lrefine\_max}-3, derefine. This criterion aims to limit the resolution of the ejecta at angles where a breakout has occurred before $t_{\rm{breakout,max}}$.
\item If $g_{\rm{fac}}<98$, refine. 
\item If $g_{\rm{fac,2}}<98$, do not derefine.
\item If $t < 0.1\,\rm{s}$, and a part of the block is inside the ignition zone, refine. This ensures maximum resolution throughout the entire ignition zone.
\end{enumerate}

After the burning stage concludes, marked by the transition of the TNDW into a shock wave and the cessation of \nick production, the highest level of resolution becomes unnecessary. To enhance computational efficiency, we employ a method to evaluate the final \mni at each time step through quadratic extrapolation based on its values over the preceding three time steps. If, during any time step, the current \mni surpasses $0.002\,\msol$ and differs from the extrapolated value by less than $0.5\%$, we increment a counter by one. Otherwise, we reset the counter to zero. Once this counter reaches 20 (i.e., after 20 consecutive time steps), we decrease the maximum resolution by one refinement level. This reduction in resolution occurs only once during the simulation and is not a repetitive process.


\section{Sensitivity to the numerical setup}
\label{sec:numericsens}

In this Appendix, we study the sensitivity of our results to the numerical setup. In Section~\ref{sec:oned}, we use 1D simulations to study the effect of numerical parameters whose impact is not related to the dimension of the simulation. In Section~\ref{sec:twod}, we study the effect of the parameters that are unique to the 2D dimensionality of the calculations. The influence of all analysed parameters is relatively small, suggesting that our numerical setup is adequate for comparing the DDM calculations with the observed $ t_0-\mnii $ relation.

\subsection{1D calculations}
\label{sec:oned}

\begin{figure}
    \includegraphics[width=0.99\columnwidth]{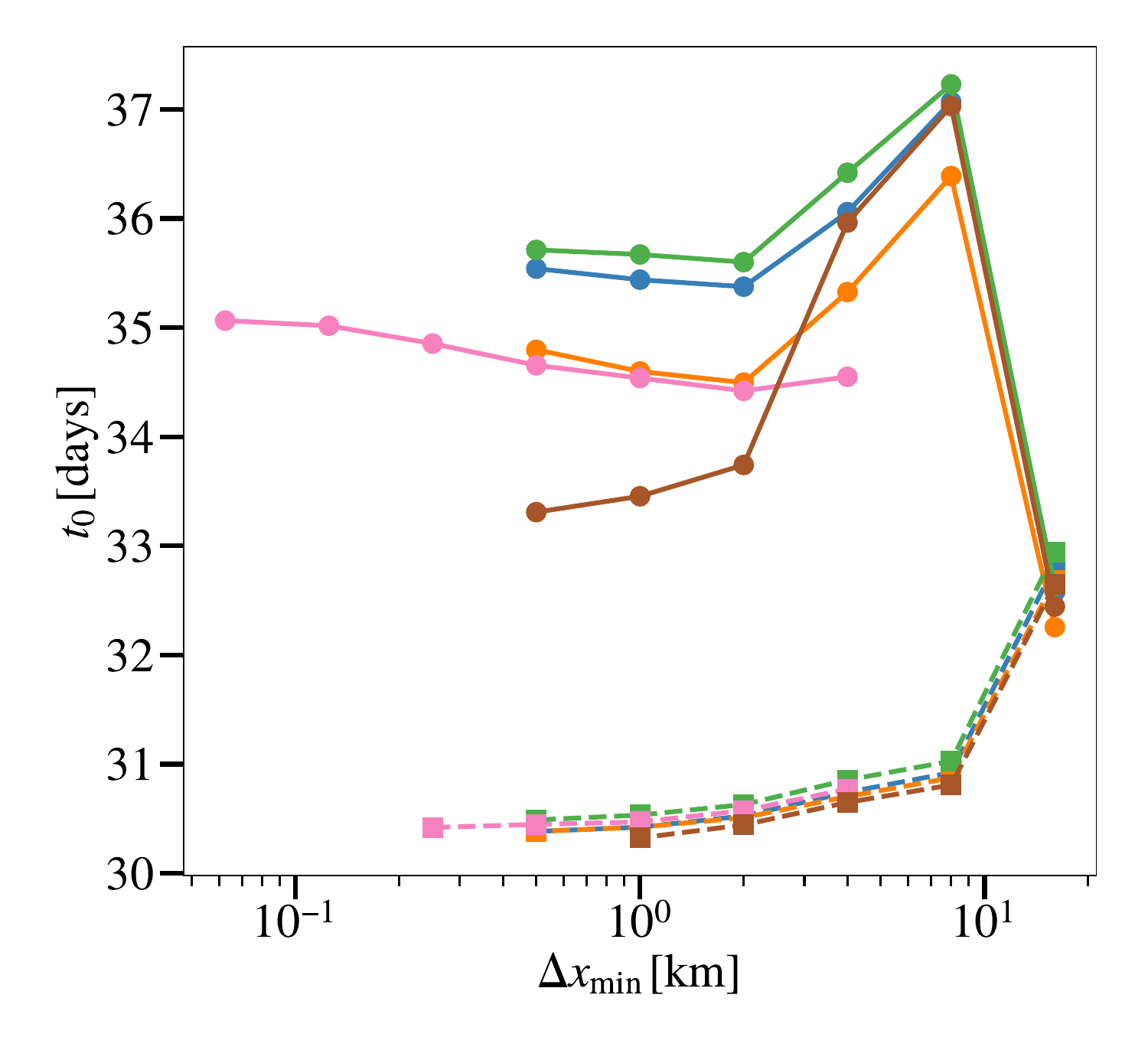}{(a)}
    \includegraphics[width=0.99\columnwidth]{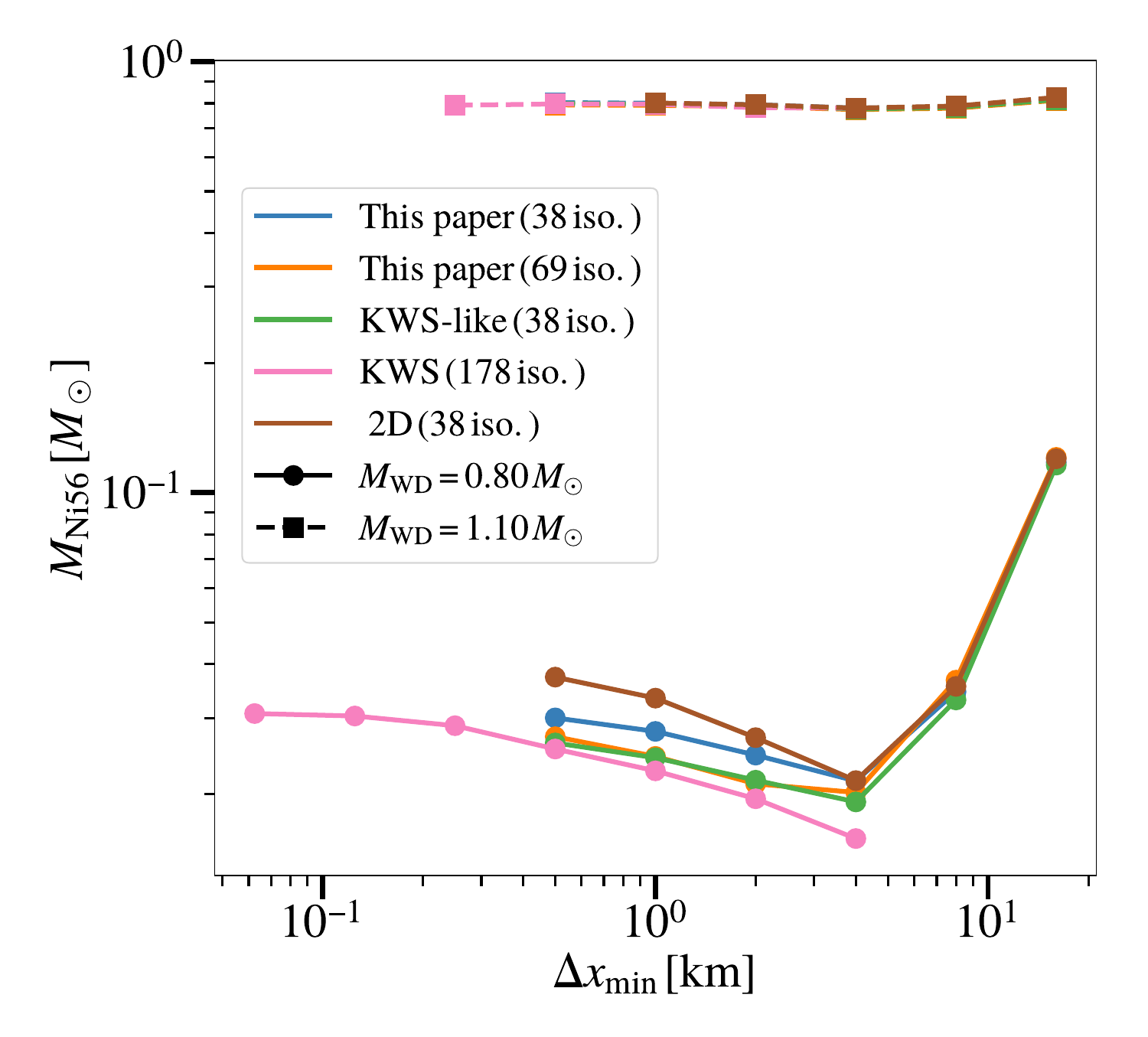}{(b)}
    \caption{The obtained $t_0$ (panel a) and \mni (panel b) as a function of resolution. Blue and orange lines: 1D calculations of our setup with 38 and 69 isotopes, respectively. Green line: calculations based on the setup of \refKWS with 38 isotopes. Pink line: results from \refKWS. Solid lines represent $\mwd=0.8\,\msol$, and dashed lines represent $\mwd=1.1\,\msol$. The results from the 69-isotope list agree well with the default results of \refKWSs, demonstrating both the accuracy of the 69-isotope list and that our more efficient scheme does not significantly compromise simulation accuracy. Our approach aligns closely with the \refKWS scheme when both use the 38-isotope list, with minor deviations compared to more detailed isotope lists. The convergence characteristics of our method and the \refKWS scheme are comparable. Brown line: 2D central ignition for $\mwd=0.8\,\msol$. The 2D results diverge from the 1D results when the resolution surpasses $\Delta x_{\rm{min}}=2\,\rm{km}$, attributed to the emergence of 2D instabilities in our simulations (see text).}
    \label{fig:1dconv}.
\end{figure}

Our numerical setup, as outlined in Section~\ref{sec:setup}, differs in several aspects from the configuration utilized in the 1D investigation by \refKWSs. These distinctions primarily stem from the need to optimize computational resources more effectively in our 2D simulations. Fig.~\ref{fig:1dconv} presents a comparison between 1D computations of central ignitions for $\mwd=0.8\,\msol$ and $\mwd=1.1\,\msol$, performed using our method (with 38 and 69-isotope lists; employing a maximum resolution of $\Delta x_{\rm{min}}\ge0.5\,\rm{km}$)  and the method of \refKWS (utilizing their default 178-isotope list; we also include a series of simulations using our 38-isotopes list). The results from the 69-isotope list calculations agree well with the default results of \refKWSs, demonstrating both the accuracy of the 69-isotope list and the fact that our more efficient scheme does not significantly compromise the accuracy of the simulation. Our approach aligns closely with the \refKWS scheme when both are computed with the 38-isotope list, although there are minor deviations compared to the more detailed isotope lists. These deviations are addressed in Section~\ref{sec:exconv} and are insignificant to our objectives. The convergence characteristics of our method and the \refKWS scheme are comparable.

\begin{figure}
\includegraphics[width=\columnwidth]{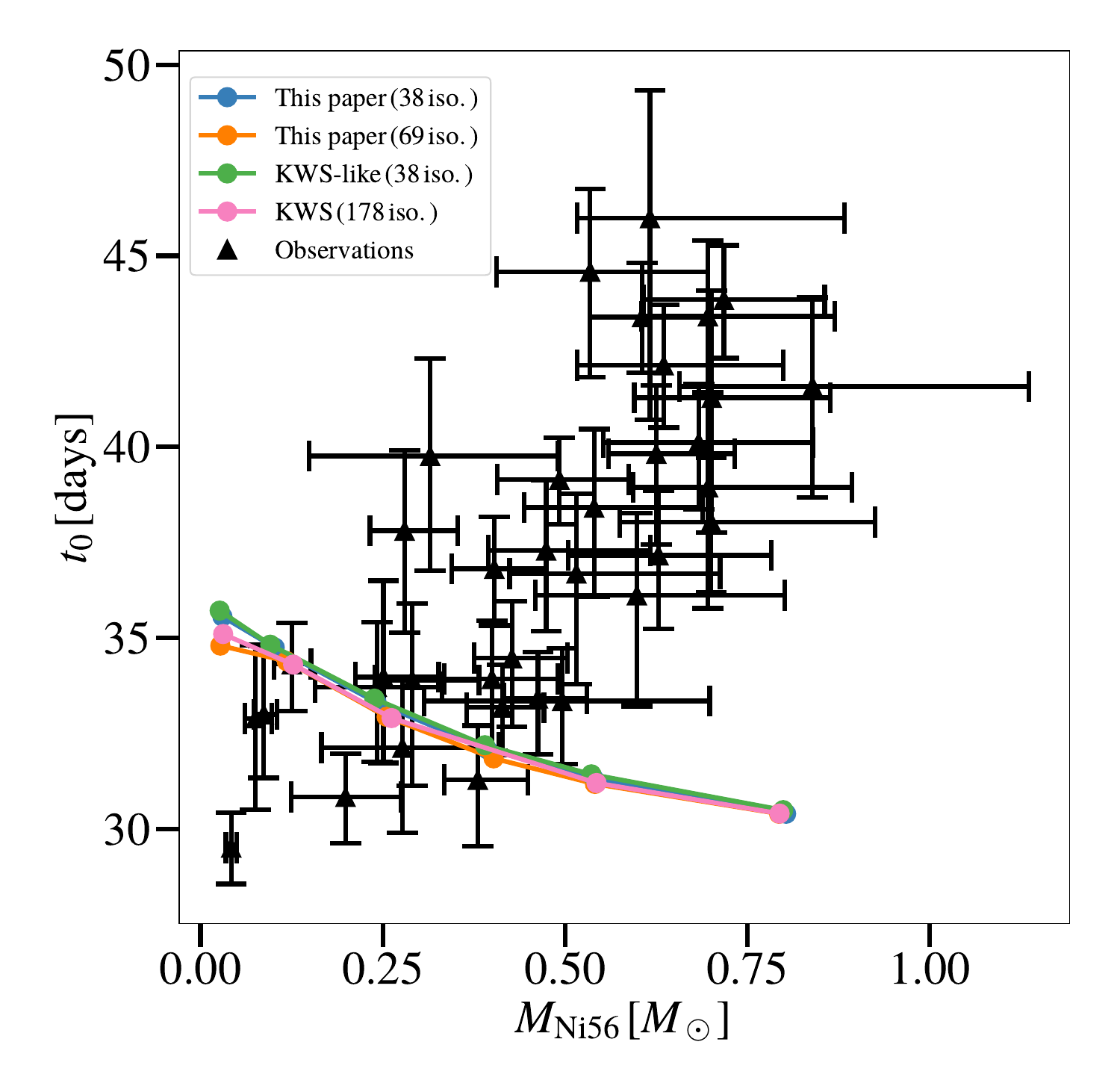}
\caption{The $t_0-M_{\rm{Mi56}}$ relation. Black circles: the observed SNe Ia sample \protect\citep{sharon_-ray_2020,sharon_personal_2023}. Each line represents the converged outcome (with a resolution of $\Delta x_{\rm{min}}=0.5\,\rm{km}$) from 1D configuration for $\mwd=0.8,0.85,0.9,0.95,1.0,1.1\,\msol$ (left to right) with the same colour scheme as in Fig.~\ref{fig:1dconv}. The disparities between our method and the approach of \refKWSs, as well as the differences between the various isotope lists, are minimal, facilitating a meaningful comparison of the calculations with the observed $t_0-$\mni relation.}
\label{fig:1dcomp}
\end{figure}

Similar trends are observed for other WD masses. Fig.~\ref{fig:1dcomp} showcases the converged outcomes (with a resolution of $\Delta x_{\rm{min}}=0.5\,\rm{km}$) from our 1D configuration for $\mwd=0.8,0.85,0.9,0.95,1.0,1.1\,\msol$ in the $t_0-$\mni plane alongside the observational data. As depicted in the Figure, the disparities between our method and the approach of \refKWSs, as well as the differences between the various isotope lists, are minimal, facilitating a meaningful comparison of the calculations with the observed $t_0-$\mni relation.

\subsection{2D calculations}
\label{sec:twod}

Certain parameters in our numerical setup may influence our 2D results in ways that do not have direct analogues in 1D simulations or only have minimal effects on 1D computations. To evaluate the sensitivity of these factors in our 2D simulations, we performed a series of calculations using the configurations outlined in Section~\ref{sec:example}, focusing on the case of $\mwd=0.8\,\msol,\zig=0.5$, while maintaining a resolution of $\Delta x_{\rm{min}} = 2\,\rm{km}$. The parameters explored in these tests are as follows:

\begin{enumerate}
    \item \texttt{cfl} - Courant-Friedrichs-Lewy factor for time-step criteria.
    \item Density and temperature thresholds for various program modules:
    \begin{enumerate}
        \item \texttt{smlrho} - minimal density for hydrodynamics.
        \item \texttt{smallt} - minimal temperature for the equation of state.
        \item \texttt{nuclearDensMin} and  \texttt{nuclearTempMin} - density and temperature thresholds for nuclear burning.
        \item \texttt{amb\_dens} and \texttt{amb\_temp} - density and temperature of the ambient material external to the WD \citep[often referred to as "fluff";][]{boos_multidimensional_2021}.
    \end{enumerate}  
    \item \texttt{mpole\_Lmax} - the maximum angular moment used in the multipole Poisson solver. 
    \item \texttt{eintSwitch} - the upper limit of the internal-to-kinetic energy ratio for updating the pressure based on internal energy. 
\end{enumerate}

The effects of these parameters are summarized in Table~\ref{tab:convtab}. Most parameters show minimal influence, typically altering the results by less than 2\%. The notable exception is the \texttt{cfl} parameter, which affects the \nick mass by approximately 11\% when halved. However, this remains small relative to the 20\% convergence error associated with $\Delta x_{\rm{min}} = 2\,\rm{km}$, as shown in Fig.~\ref{fig:flim}. When the \texttt{cfl} is similarly reduced in the $\Delta x_{\rm{min}} = 1\,\rm{km}$ case, the changes are significantly smaller, with a 2\% variation in $\mnii$ and a 0.3\% change in $t_0$.


\begin{table}
	\centering
	\caption{Simulation parameters examined in sensitivity tests for the case of $\mwd=0.8\,\msol,\zig=0.5$ with $\Delta x_{\rm{min}}=2\,\rm{km}$.\label{tab:convtab}}
	\label{tab:setup_param}
	\begin{tabular}{lccccr} 
		\hline
		Parameter & Default & More accurate & $\Delta\mnii$ & $\Delta t_0$ \\
		\hline
		\texttt{cfl} & 0.8 & 0.4 & 11 per cent& 0.03 per cent\\
            \hline
		\texttt{smlrho} & $10^{-3}$ & $10^{-4}$\\
		\texttt{amb\_dens} & $10^{-2}$ & $10^{-3}$\\
  		\texttt{nuclearDensMin} & $10^{5}$ & $10^{4}$ & 2 per cent& 0.7 per cent\\
		\texttt{smallt} & $10^{6}$ & $10^{5}$\\
		\texttt{amb\_temp} & $10^{7}$ & $10^{6}$\\
  		\texttt{nuclearTempMin} & $10^{8}$ & $10^{7}$\\
            \hline
            \texttt{mpole\_Lmax} & $16$ & $32$ & 0.9 per cent& 0.2 per cent\\
            \hline
            \texttt{eintSwitch} & $0.1$ & $1$ & 1 per cent& 0.3 per cent\\
		\hline
	\end{tabular}
\end{table}

An artefact in our scheme arises when early ejected material may be excluded from the calculation if it reaches a discrete angular segment that has not been breached. In such instances, we enforce the state of the regions outside the TNDW front to be the initial state of the calculation, potentially overriding exterior regions containing part of the ejecta. This issue seems to occur primarily at our most extreme ignition location of $\zig=0.75$. However, since the density of the ejecta in these regions, which could be removed, is of the order of $\mysim10\,\rm{g/cm^3}$, the impact of this artefact should be negligible. To assess this effect, we calculated the case of $\mwd=1\,\msol$ and $\zig=0.75$ (with $ \Delta x_{\rm{min}}=2\,\rm{km} $) using 16 angular segments instead of our default 50 segments. Furthermore, we conducted calculations for this case with a more refined angular treatment that considers this part of the ejecta without enforcing the initial state on its angular segment. As anticipated, all these modifications have a subpercent effect on the results.


\section{TNDW instability}
\label{sec:detinstab}

Fig.~\ref{fig:main_plot} highlights a significant observation: central ignition exhibits differences between the 1D and 2D models, especially noticeable in low-mass WDs. This contrast becomes apparent in Fig.~\ref{fig:1dconv}, where we observe that the outcomes of the 2D central ignition for $\mwd=0.8\,\msol$ diverge from the 1D simulation when the resolution surpasses $\Delta x_{\rm{min}}=2\,\rm{km}$. We attribute this variance to the emergence of 2D instabilities in our simulations.

\begin{figure*}
\includegraphics[height=7cm]{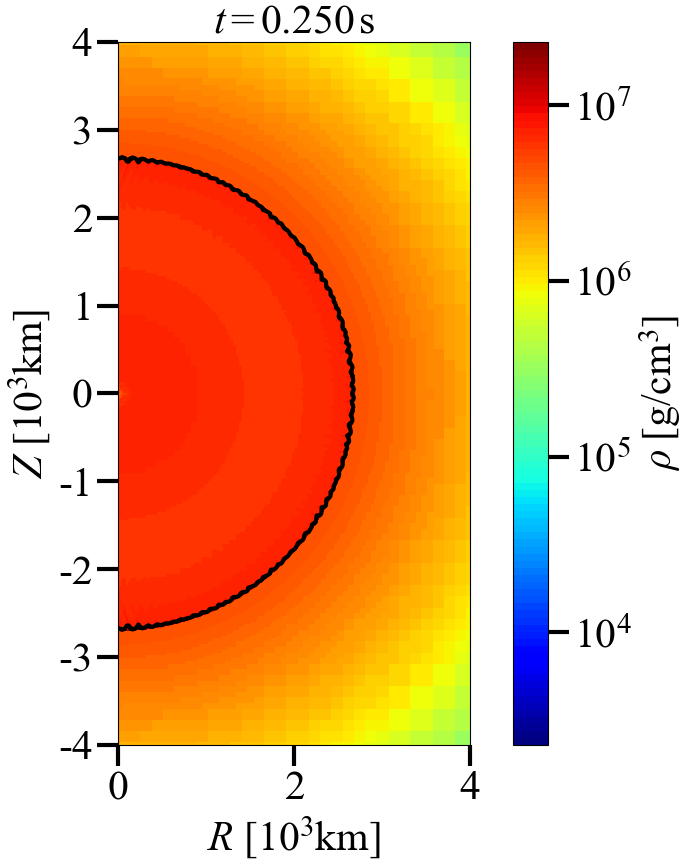}{(a)}
\includegraphics[height=7cm]{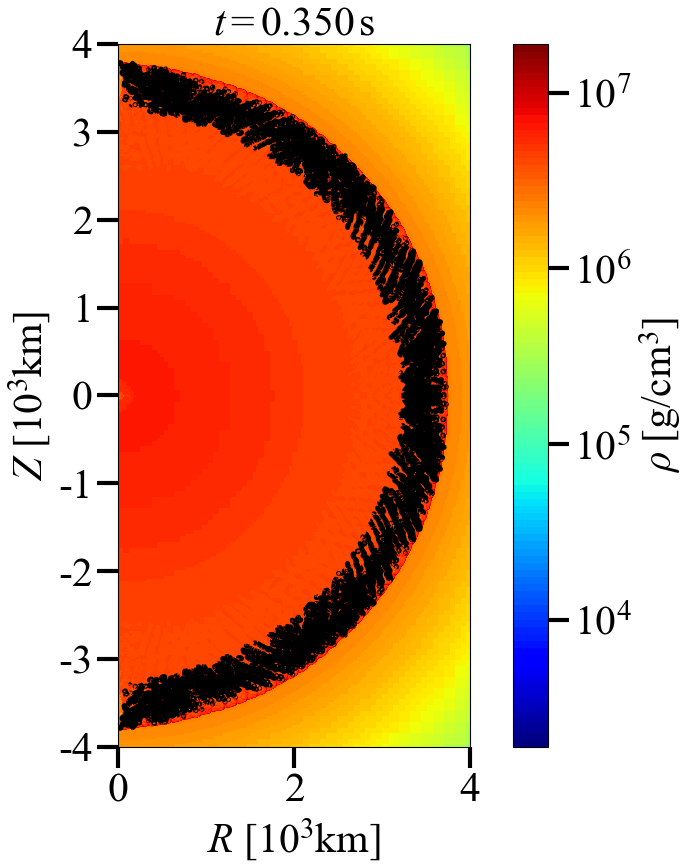}{(b)} \\
\includegraphics[height=7cm]{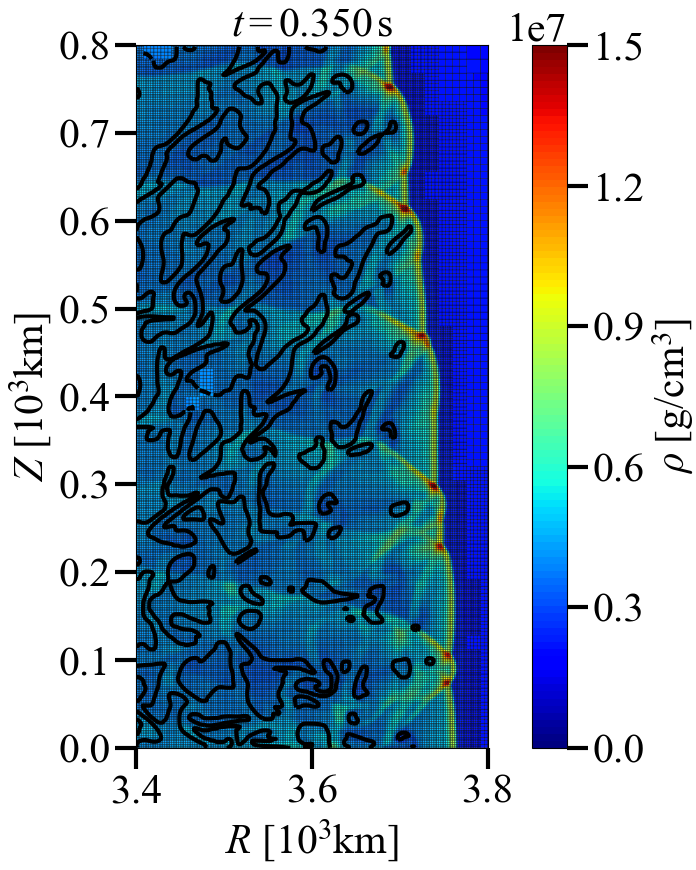}{(c)}
\includegraphics[height=7cm]{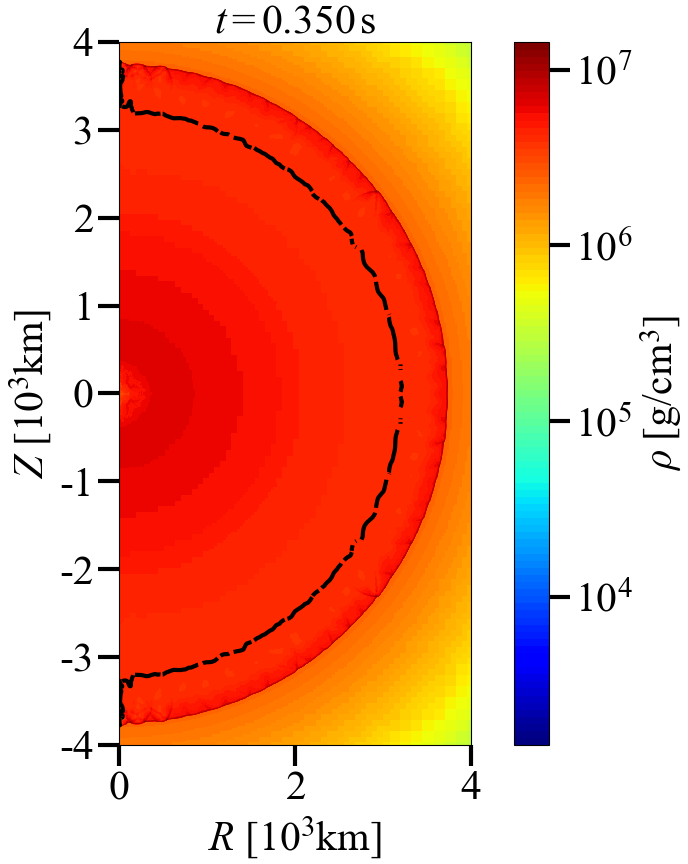}{(d)}
\caption{Density colour maps of $\mwd=0.8\msol$ 2D calculations with central ignition. Panels (a)-(c) display a calculation with a resolution of $\Delta x_{\rm{min}}=2\,\rm{km}$ at various times. At $t=0.25\,\rm{s}$ [panel (a)], the TNDW front appears spherical; however, at later times [$t=0.35\,\rm{s}$; panel (b)], a 2D instability emerges in the $^{56}$Ni front. Panel (c) presents a magnified view of panel (b), showcasing the computational mesh. The features of the 2D instability surpass the scale of individual computational cells and, thus, are fully resolved. Panel (d) illustrates that at a lower resolution of $\Delta x_{\rm{min}}=4\,\rm{km}$, the $^{56}$Ni front appears much more symmetric at the same time. This implies that a minimum resolution ($\Delta x_{\rm{min}}=2\,\rm{km}$ in our case) is required to capture the TNDW instability.}
\label{fig:sympic}
\end{figure*}

Panels (a) and (b) of Fig.~\ref{fig:sympic} display the colour density map of 2D calculation with a maximum resolution of $\Delta x_{\rm{min}}=2\,\rm{km}$ at various times. At $t=0.25\,\rm{s}$, the TNDW front appears spherical; however, at later times ($t=0.35\,\rm{s}$), a 2D instability emerges in the $^{56}$Ni front. Panel (c) in the same figure presents a magnified view of panel (b), showcasing the computational mesh. The features of the 2D instability surpass the scale of individual computational cells and, thus, are fully resolved. These characteristics bear similarity to those described previously in the literature \citep{timmes_cellular_2000,Parete2012,Papatheodore2014}, albeit on a larger scale of kilometers instead of centimeters. Panel (d) of Fig.~\ref{fig:sympic} illustrates that at a lower resolution of $\Delta x_{\rm{min}}=4\,\rm{km}$, the $^{56}$Ni front appears much more symmetric at the same time. This implies that a minimum resolution ($\Delta x_{\rm{min}}=2\,\rm{km}$ in our case) is required to capture the TNDW instability. Conversely, the size of the instability features is resolution-dependent, with more features observed at a $\Delta x_{\rm{min}}=1\,\rm{km}$ resolution. 

When comparing \mnis, the 2D calculation with a resolution of $\Delta x_{\rm{min}}=4\,\rm{km}$ for $\mwd=0.8\,\msol$ closely resembles its 1D counterpart (differences less than $1\,\%$). In contrast, the $\Delta x_{\rm{min}}=2\,\rm{km}$  ($\Delta x_{\rm{min}}=1\,\rm{km}$ ) resolution deviates by approximately $\mysim10\,\%$ ($\mysim20\,\%$) from its 1D equivalent. A similar trend is observed for $t_0$: the 2D calculation at a resolution of $\Delta x_{\rm{min}}=2\,\rm{km}$  ($\Delta x_{\rm{min}}=1\,\rm{km}$ ) resulting in $t_0\myapprox33.8\,\rm{days}$ ($t_0\myapprox33.5\,\rm{days}$), while the 1D simulations (with resolutions ranging from $\Delta x_{\rm{min}}=1\,\rm{km}$ to $\Delta x_{\rm{min}}=4\,\rm{km}$) and the 2D simulation with a resolution of $\Delta x_{\rm{min}}=4\,\rm{km}$ have $t_0\myapprox35.3\pm0.1\,\rm{days}$. 

Given that the instability, especially in low-mass WDs, influences our results, it raises the question of whether this effect is a numerical artefact inherent to our scheme. The TNDW instability may impact results when the instability scale is comparable to the pressure gradient scale of the WD \citep{khokhlov_stability_1993,timmes_cellular_2000}. While our results demonstrate convergence in spatial resolution (which scales with instability features), we assert that this effect is accurately captured, although we cannot rule out the possibility of additional influences from low-scale instabilities, such as cm-scale instabilities.

The choice of $\flim=0.1$ might introduce a potential bias to the results, as the description of instability could be sensitive to this parameter. As depicted in Fig.~\ref{fig:flim}, the impact of $\flim$ is more pronounced in scenarios involving low-mass WDs, where the instability plays a more significant role. However, even in these instances, the effect on the results appears to be minimal, on the order of a few percentages at most. This magnitude is smaller compared to the influence of $\mwd$ or $\zig$ on the outcomes, thus it does not alter the conclusions of this study regarding the tension between our simulation and the observation of the $ t_0-\mnii $ relation. A more thorough understanding of how the resolution and the burning limiter affect the numerical description of TNDW instabilities will be pursued in future research.

\label{lastpage}
\end{document}